\documentclass[format=manuscript, screen, review=false]{acmart}

\AtBeginDocument{%
  }

\setcopyright{acmlicensed}
\copyrightyear{2025}
\acmYear{2025}
\acmDOI{XXXXXXX.XXXXXXX}

\acmJournal{TOPLAS}
\acmVolume{62}
\acmNumber{4}
\acmArticle{111}
\acmMonth{8}

\usepackage{wrapfig}
\usepackage[linesnumbered,boxed]{algorithm2e}
\usepackage{mathtools}
\usepackage{multirow}
\usepackage{caption}
\usepackage{subcaption}
\usepackage{amsthm}
\newtheorem*{c17}{COROLLARY 17}
\newtheorem*{t16}{THEOREM 16}
\newtheorem*{c15}{COROLLARY 15}
\newtheorem*{t2}{THEOREM 2}
\newtheorem*{l1}{LEMMA 1}
\newtheorem*{t19}{THEOREM 19}

\begin{document}

\title{Futureproof Static Memory Planning}

\author{Christos P. Lamprakos}
\email{cplamprakos@microlab.ntua.gr}
\orcid{0000-0002-3370-857X}
\affiliation{%
  \institution{National Technical University of Athens}
  \country{Greece}
}
\affiliation{%
  \institution{KU Leuven}
  \country{Belgium}
}

\author{Panagiotis Xanthopoulos}
\email{panos0511@gmail.com}
\affiliation{%
  \institution{National Technical University of Athens}
  \country{Greece}
}

\author{Manolis Katsaragakis}
\email{mkatsaragakis@microlab.ntua.gr}
\orcid{0000-0001-8116-3503}
\affiliation{%
  \institution{National Technical University of Athens}
  \country{Greece}
}
\affiliation{%
  \institution{KU Leuven}
  \country{Belgium}
}

\author{Sotirios Xydis}
\email{sxydis@microlab.ntua.gr}
\orcid{0000-0003-3151-2730}
\author{Dimitrios Soudris}
\email{dsoudris@microlab.ntua.gr}
\orcid{0000-0002-6930-6847}
\affiliation{%
  \institution{National Technical University of Athens}
  \country{Greece}
}

\author{Francky Catthoor}
\email{francky.catthoor@imec.be}
\orcid{0000-0002-3599-8515}
\affiliation{%
  \institution{National Technical University of Athens}
  \country{Greece}
}
\affiliation{%
  \institution{KU Leuven}
  \country{Belgium}
}

\renewcommand{\shortauthors}{Lamprakos et al.}

\begin{abstract}
The NP-complete combinatorial optimization task of assigning offsets to a set of buffers with known sizes and lifetimes so as to minimize total memory usage is called dynamic storage allocation (DSA). Existing DSA implementations bypass the theoretical state-of-the-art algorithms in favor of either fast but wasteful heuristics, or memory-efficient approaches that do not scale beyond one thousand buffers. The ``AI memory wall'', combined with deep neural networks' static architecture, has reignited interest in DSA. We present \texttt{idealloc}, a low-fragmentation, high-performance DSA implementation designed for \textit{million}-buffer instances. Evaluated on a novel suite of particularly hard benchmarks from several domains, \texttt{idealloc} ranks first against four production implementations in terms of a joint effectiveness/robustness criterion.
\end{abstract}




\maketitle

\section{Introduction}
\label{sec:intro}
Deep learning is causing significant shifts in professional and civilian life. Several technical challenges, however, remain open. For instance, there is a profound asymmetry between progress in compute capability and memory capacity/bandwidth~\cite{ai_memory_wall}. This so-called ``AI memory wall'' has sparked substantial research and engineering efforts targeting the memory effectiveness of deep learning. The particular line of work in scope for this paper deals with assigning offsets to a set of buffers with known sizes and lifetimes in order to pack them in as small an address space as possible~\cite{minimalloc,telamalloc,somas,triton,footprint,ismm2,levental2022memory,pisarchyk2020efficient,memoLLM}. In deep learning such problems appear thanks to (i) neural networks' static architecture and (ii) hardware accelerators' physical memory contiguity.

Nevertheless, beyond providing motivation for what shall be presented, deep learning is not of the essence here. The problem is old and well-studied~\cite{garey1979computers,kierstead_1,KIERSTEAD1991231,gergov_1,gergov_2,buchsbaum}. It is known as \textit{dynamic storage allocation} (DSA), a variation of two-dimensional bin packing. DSA has been proven NP-complete.

\subsection{Against a Common Misunderstanding}
Despite its name, DSA is a \textit{static} problem, in the sense of having available all the information that it needs from the outset. ``Dynamic storage allocation'' has also been used for the dynamic variant (what \texttt{malloc} implementations deal with)~\cite{wilson,robson_1,robson_2}, causing considerable confusion. We shall be using ``DSA'', ``memory planning'', ``static offset assignment'' and ``static memory allocation'' interchangeably in this text. In a similar vein, we will be referring to DSA implementations, i.e., programs solving DSA instances, as ``allocators''. Dynamic non-moving virtual memory allocators such as GNU's \texttt{malloc} are out of scope---we use ``OS allocators'' in the few times that we must mention them.

\subsection{Motivation and Related Work}
We are concerned with \textit{real-world implementations} of DSA, their \textit{effectiveness}, \textit{efficiency} and \textit{robustness} in the face of arbitrarily large inputs. Our founding assumption is that sooner or later, in deep learning or elsewhere, DSA instances comprising \textit{millions} of buffers will emerge. For instance, large language models are already pushing compiler engineers to come up with ever more aggressive optimizations, yielding complex and massive memory allocation patterns in return~\cite{ismm2,gmlake}. Another example is the Linux user applications domain, where \texttt{malloc} traces are used for off-line analysis and/or optimization~\cite{beyondrss,halo,idealloc_1,learningmalloc,maphea,navasca}.

Our main observation after surveying the state-of-the-art (SOTA) was that allocators are bypassing the algorithms published in the DSA literature in favor of schemes that are simpler to implement. Alternatives can be sorted in two broad categories: heuristics~\cite{pisarchyk2020efficient,somas,sekiyama2018profileguidedmemoryoptimizationdeep,levental2022memory} and isomorphisms~\cite{minimalloc,telamalloc,deeploy}, e.g., integer linear programming, machine learning regression, simulated annealing, and hill-climb optimization. We ask what costs accompany circumventing the decades-old literature around an NP-complete problem for which one seeks a practical, general solution. The only way to find out \textit{if} such costs exist would be to build an allocator informed by that literature, and then evaluate it rigorously against the SOTA. Hence \texttt{idealloc}, the allocator at this paper's center, was born. In terms of the heuristics/isomorphisms dichotomy, it is a \textit{stochastic bootstrapped heuristic}.

A second observation was that apart from the micro-benchmarks published by the authors of \texttt{minimalloc}, a SOTA allocator~\cite{minimalloc}, no DSA benchmark suites exist. We thus formed a novel set of benchmarks ranging from hundred- to half-a-million buffers and used it, along with the aforementioned micro-benchmarks, for evaluation. From a strict effectiveness-only perspective \texttt{idealloc} rarely beats all of its competition, comprising \texttt{minimalloc} and three other production allocators. But from a robustness and efficiency perspective that same competition (with one exception) rarely manages to even produce a solution in reasonable time. Under a joint ranking criterion incorporating both perspectives \texttt{idealloc} achieves top score.

\subsection{Contributions}
Along the course of designing, developing and testing \texttt{idealloc}, we gathered a multifaceted set of insights. On the algorithmic front, we identified and fixed several blind spots of the original theorems, published by Buchsbaum et al. in 2003~\cite{buchsbaum}~\footnote{We have exchanged emails with the algorithm's original authors, who have validated that (i) transition from theory to practice always involves trickiness and (ii) there are no other known implementations of their work.}. We also devised a second set of algorithms, related not to the DSA core itself, but to forming a scalable infrastructure around it. On the benchmarks front, we collected a novel suite of challenging, large-scale inputs from domains such as Linux databases, parallel training of deep learning models, and distributed inference. 

All in all, our contributions are:

\begin{enumerate}
    \item \texttt{idealloc}, a DSA implementation designed to handle inputs of arbitrary size and complexity
    \item crucial theoretical extensions to the algorithms on which \texttt{idealloc} is based
    \item various insights and techniques of general applicability to future DSA design tasks
    \item the first rigorous evaluation of the DSA SOTA
\end{enumerate}

Section \ref{sec:DSA} provides background knowledge on DSA. Section \ref{sec:coreba} describes the core algorithm powering \texttt{idealloc}. The design of our allocator is exposed in detail in Section \ref{sec:design}, and the experiments conducted for evaluating it are reported in Section \ref{sec:xps}. Section \ref{sec:disc} discusses limitations and ideas for future work, and Section \ref{sec:end} concludes our exposition.

\section{Dynamic Storage Allocation}
\label{sec:DSA}
Rectangle packing~\cite{rectpack} is the combinatorial optimization problem of placing rectangles of various widths and heights into arrangements where (i) no two rectangles overlap and (ii) the arrangement's enclosing rectangle has minimum height. Rectangles may move in two degrees of freedom (vertically or horizontally). This problem is NP-complete.

\begin{wrapfigure}[18]{r}{0.5\textwidth}
    \begin{center}
        \includegraphics[width=0.48\textwidth]{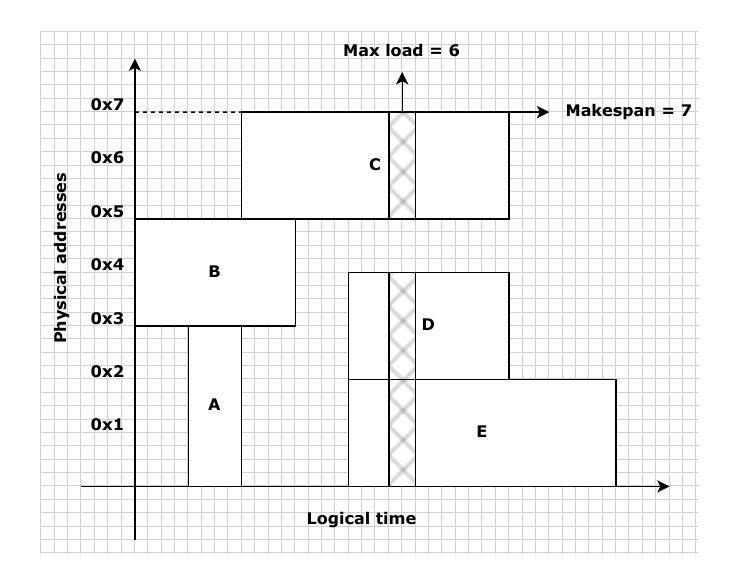}
    \end{center}
    \caption{A more detailed illustration of the dynamic storage allocation (DSA) problem. This instance comprises five buffers and a (suboptimal) solution, i.e., offset assignment to each of the buffers, is depicted.}
    \label{fig:2dbptut}
\end{wrapfigure}

DSA is a constrained variation of rectangle packing. It owes its name to the interpretation of one dimension as available address space, and the other as time. Each rectangle encodes a pair of requests for the allocation and deallocation of some specific amount of memory at specific points in time. Allocators have no power over the timing of incoming requests, so the only degree of freedom they have is the spatial one. DSA is NP-complete in the general case of non-uniform request sizes. A toy illustration comprising three rectangles is shown at Figure \ref{fig:2dbptut}. If the horizontal axis represents time, the allocator may move rectangles vertically.

A DSA \textbf{input} comprises buffers defined as $(h, t_s, t_e)$ tuples, where $h$ stands for buffer size. All of the data involved are \textbf{discrete}, more precisely non-negative integers. A buffer is \textbf{live} in the open interval $(t_s, t_e)$. We refer to $(t_s, t_e)$ as the buffer's \textbf{lifetime}. We refer to $t_s$ and $t_e$ as \textbf{allocation time} and \textbf{de-allocation time} respectively. A buffer's \textbf{lifespan}, i.e., the size of its lifetime, i.e., the total number of time units at which the buffer is live, is computed as below:

\begin{equation}
    l = t_e - t_s - 1
\end{equation}

Two buffers \textbf{overlap} if their respective lifetimes overlap. An input's \textbf{load} at moment $t$ is the size sum of all buffers live at $t$. We refer to the maximum load measured across all $t$ as \textbf{max load} ($L$). In Figure \ref{fig:2dbptut}, the max load is the length of the cross-hatched stripe. The small gap between the two pieces does not contribute to it because it does not belong to any buffer, it's just unused space. By \textbf{placement} we mean annotating an input's buffers with valid offsets. A placement's \textbf{makespan} or \textbf{max memory usage} ($M$) is the address space size needed to fit all buffers. \textbf{Fragmentation} ($F$) is the difference between an input's max load and the actual makespan of some placement ($7-6=1$ byte in Figure \ref{fig:2dbptut}):

\begin{equation}
    F = M - L
\end{equation}

\IncMargin{0.5em}
\begin{wrapfigure}[33]{r}{0.52\textwidth}
    \begin{minipage}{0.52\textwidth}
        \begin{algorithm}[H]
            \SetKwProg{Fn}{Function}{}{end}
            \SetKwFunction{IGC}{IntervalGraphColoring}
            \SetKwFunction{V}{OffsetsValid}
            \SetKwFunction{GE}{GetEvents}
            \SetKwFunction{A}{IsAlloca}
            \SetKwFunction{F}{IsFree}
            \SetKwInOut{Input}{input}
            \SetKwInOut{Output}{output}
            \Fn{\IGC{$B$}} {
                \tcp{A set of same-size buffers.}
                \Input{$~B = \{~b~|~b=(h,t_s,t_e)~\}$}
                \tcp{A valid buffer-offset mapping.}
                \Output{$~O = \{~o~|~o \in \mathbb{N} : \V{B, O}~\}$}
                \BlankLine
                
                $O\leftarrow$ \texttt{HashMap.new();}\\
                \tcp{Buffer-row mapping.}
                \texttt{live} $\leftarrow$ \texttt{HashMap.new();}\\
                \tcp{Returns lowest free row upon pop().}
                \texttt{free} $\leftarrow$ \texttt{PriorityQueue.new();}\\
                \tcp{To be used if no free row exists.}
                \texttt{next\_row} $\leftarrow$ \texttt{0;}\\
                \tcp{(De-)allocations priority queue.}
                \texttt{evts} $\leftarrow$ \GE{$B$}\texttt{;}\\
                \BlankLine
                \tcp{Each .pop() spawns an ``e''.}
                \While{\texttt{evts.pop()}}{
                    \uIf{\A{\texttt{e}}}{
                        \uIf{\texttt{free.empty()}}{
                            \texttt{offset} $\leftarrow$ \texttt{next\_row;}\\
                            \texttt{next\_row += 1;}
                        } \Else{
                            \texttt{offset} $\leftarrow$ \texttt{free.pop();}\\
                        }
                        \texttt{live.insert((e.buff, offset));}\\
                        $O$\texttt{.insert((e.buff, offset));}\\
                    } \Else{
                        \texttt{freed\_row} $\leftarrow$ \texttt{live.remove(e.buff);}\\
                        \texttt{free.push(freed\_row);}\\
                    }
                }
                \BlankLine
                \KwRet{O}\texttt{;}
            }
        \end{algorithm}        
    \end{minipage}
    \caption{Interval Graph Coloring.}
    \label{algo:igc}
\end{wrapfigure}

The NP-completeness of DSA has led researchers toward approximation algorithms. The quality of each algorithm is expressed as upper bounds for fragmentation. For instance, a 6-approximation algorithm guarantees that it will never produce a makespan six times bigger than the max load. The current SOTA in DSA is a $(2+\epsilon)$-approximation algorithm by Buchsbaum et al~\cite{buchsbaum}. $\epsilon$ is described as a ``sufficiently small'' real number and is input-dependent.

\subsection{Elementary Cases}
\label{sec:elem}
There are certain instances of the problem which can be solved optimally, i.e., with zero fragmentation. One can recognize such instances in linear time. It suffices to traverse the input once, and check if (i) any overlapping buffers, or (ii) more than one buffer sizes exist. If no buffers overlap, they can all be placed at offset zero.

If all buffers share the same size, the problem is reduced to meeting room scheduling and can be solved with greedy interval graph coloring (IGC). Since we shall make use of IGC later, we remind it to the reader via Figure \ref{algo:igc}.

\subsection{Heuristics}
\label{sec:heur}
In Section \ref{sec:intro} we claimed that existing DSA implementations can be categorized as either heuristics or isomorphisms. While ``isomorphisms'' is a deliberately vague term, by ``heuristics'' we mean a specific family of solutions.

\begin{wrapfigure}[30]{r}{0.52\textwidth}
    \begin{minipage}{0.52\textwidth}
        \begin{algorithm}[H]
            \SetKwProg{Fn}{Function}{}{end}
            \SetKwFunction{FF}{FirstFit}
            \SetKwFunction{V}{OffsetsValid}
            \SetKwFunction{GNA}{GetNextAddr}
            \SetKwFunction{GC}{GetConflicts}
            \SetKwFunction{FG}{Fits}
            \SetKwInOut{Input}{input}
            \SetKwInOut{Output}{output}
            \SetKwData{R}{run}
            \SetKwData{BF}{buff}
            \SetKwData{C}{conf}
            \Fn{\FF{$B$}} {
                \tcp{A sorted set of buffers.}
                \Input{$B = \{~b~|~b=(h,t_s,t_e)~\}$}
                \tcp{A valid offset-buffer mapping.}
                \Output{$O = \{~o~|~o \in \mathbb{N} : \V{B, O}~\}$}
                \BlankLine
                $O\leftarrow$ \texttt{HashMap.new();}\\
                \tcp{Each .pop() spawns a ``buff''.}
                \While{$B$\texttt{.pop()}}{
                    \tcp{For traversing the address space.}
                    \R $\leftarrow$ \texttt{0;}\\
                    \tcp{Scan placed, conflicting buffers}
                    \tcp{in ascending offset order.}
                    \For{\C in \GC{$O$\texttt{, \BF}}}{
                        \tcp{\C.offset - \R $\geq$ \BF.size}
                        \uIf{\FG{\BF, \R, \C}}{
                            \texttt{break;}
                        } \Else {
                            \tcp{\C.offset + \C.size}
                            \R $\leftarrow$ \GNA{\C}\texttt{;}
                        }
                    }
                    $O$\texttt{.insert((buff, run));}
                }
                \BlankLine
                \KwRet{$O$};
            }
        \end{algorithm}        
    \end{minipage}
    \caption{First-fit placement.}
    \label{algo:bf}
\end{wrapfigure}

In this paper, we define a heuristic as a two-phase operation comprising (i) a \textit{sorting} step and (ii) a \textit{fitting} step. In the first step, buffers are ordered according to some arbitrarily complex criterion, e.g., decreasing size, increasing allocation time, etc. Then, during the fitting step, the sorted buffers are traversed and assigned an offset in a best- or first-fit fashion. These fits differ from what the corresponding terms mean in the OS allocators context, since DSA also cares about lifetimes. By rejecting gaps lower in the address space for better-sized gaps higher up, DSA best-fit risks being unable to fill the lower gaps later because of conflicts in the temporal domain. A counterintuitive fact stemming from this is that \textit{first-fit often incurs less fragmentation than best-fit.} Figure \ref{algo:bf} describes first-fit in detail.

\section{The Boxing Algorithm by Buchsbaum et al.}
\label{sec:coreba}
The best known DSA ``algorithm'' is a 2-approximation technique published more than two decades ago~\cite{buchsbaum}. We put quotes around the term since, as will be shown in this section, we are dealing in fact with a complex system of interacting algorithms. From now on we will be referring to that original paper as ``BA''.

We have studied BA once more in the past~\cite{idealloc_1}. Our previous implementation, despite being an indispensable research milestone, carried serious weaknesses. First of all, we never published its source code. Moreover, it suffered from severe instability, e.g., yielding out-of-memory errors for two thousand buffers, but converging as it should for twenty thousand. Most importantly, \textit{it was incorrect}: BA has latent invariants which we had not discovered back then. Violating those invariants may lead to convergence, but the converged-upon output will be far from ideal. In consequence, we were getting nonsensical results where on-line algorithms were incurring less fragmentation than our off-line, supposedly SOTA allocator~\footnote{This was an ``intellectual abstract'' paper, with the focus being on the ideas instead of the experiments. The main idea was to view OS allocators as black-box DSA agents and see how they fare against a ``standard'' DSA solution.}.

This paper aims to establish an open-source reference implementation that is correct, robust and fast. The present section handles the part about correctness. We shall do a guided tour of BA, which is \texttt{idealloc}'s beating heart. We will clarify which parts of it we kept, which ones we modified and how, and what novel additions we had to make in order to bring it to life.

\subsection{Overview}
\label{sec:baover}

\begin{figure}[t!]
    \centering
    \includegraphics[width=\textwidth]{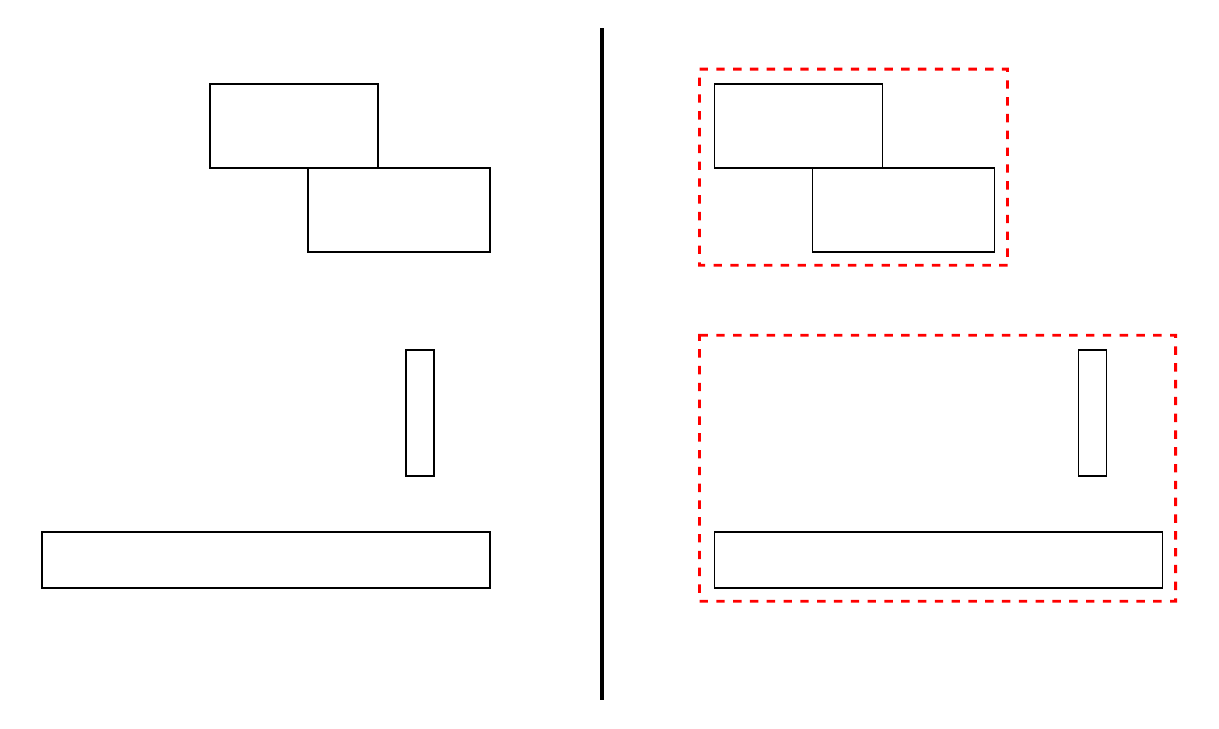}
    \caption{An illustration of BA's main idea, that is, boxing buffers into Matryoshkas. The buffers on the left have $4!=24$ possible orderings. By boxing them into two distinct groups the number of possible orderings has been reduced by a factor of $3$. In their paper, Buchsbaum et al. do not care about this reduction in complexity; they use the boxes to reason about worst-case fragmentation.}
    \label{fig:batut}
\end{figure}

The most important thing to understand about BA is that it is \textit{incomplete}. In the heuristics terminology introduced in Section \ref{sec:heur}, BA is a partial sorting step. It accepts a set of buffers as input, and yields a set of Matryoshka doll-like boxes as output. These boxes contain other boxes, and so on until some level of depth where subsets of the original input's buffers reside (see Figure \ref{fig:batut}). To keep consistent with BA's terminology, we will be referring to both the input's buffers and the output's boxes as \textbf{jobs}. The key characteristic of the outermost jobs is that they all share the same size, and as Section \ref{sec:elem} notes, they can be optimally placed with IGC. How offsets given to the top-level Matryoshkas should bleed through each boxing layer, eventually to reach the original buffers at the bottom, is not treated by BA's authors. We shall return to this question in Section \ref{sec:unbox}. For now, let us focus on the process followed to convert BA into source code.

Like any mathematics paper, BA comprises lemmas, theorems and corollaries. We will be referring to these constructs collectively as \textit{functional units} (FUs). FUs are numbered in the order that they appear in the paper: Lemma 1 is followed by Theorem 2, then comes Lemma 3 and so on.

Each FU comprises a \textit{statement}, and a \textit{proof} testifying to the correctness of the statement. The rather convenient characteristic of BA is that all of its proofs are made by construction. Every step of every proof either \textit{calculates} something (e.g., ``compute the min/max ratio of input job sizes'') or \textit{invokes} some other FU. Thus, to implement BA it suffices to view each FU as a program function, and each proof as the corresponding function body. To give a concrete example, consider Corollary 17, which we initially took to be BA's ``entry point'':

\begin{c17}
There exists a polynomial-time algorithm that takes an arbitrary set $X$ of jobs as input and produces a feasible solution to DYNAMIC STORAGE ALLOCATION on $X$ with makespan at most $(1 + O((h_{max}/L)^{1/7}))L$.
\end{c17}

\begin{proof}
    Apply Theorem 16 to $X$ with $\epsilon = (h_{max}/L)^{1/7}$.
\end{proof}

\begin{wrapfigure}[24]{r}{0.52\textwidth}
    \begin{minipage}{0.52\textwidth}
        \begin{figure}[H]
            \centering
            \includegraphics[width=0.99\textwidth]{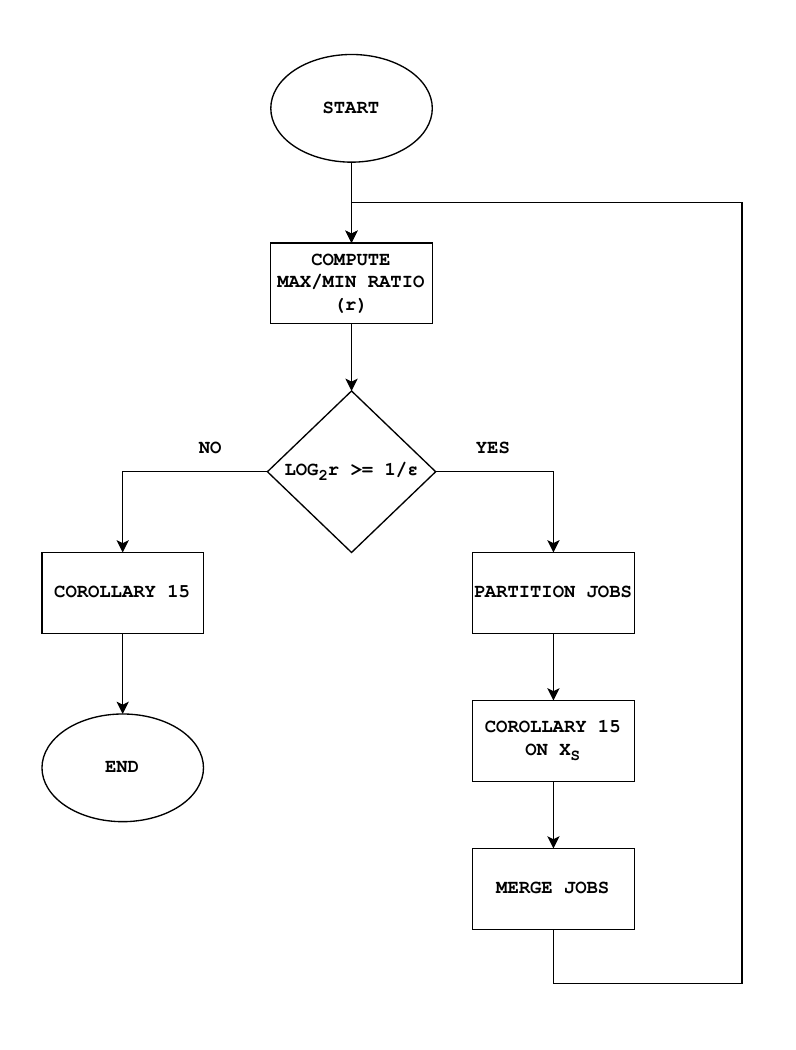}
        \end{figure}       
    \end{minipage}
    \caption{T16 flow diagram.}
    \label{fig:t16}
\end{wrapfigure}

Recall from Section \ref{sec:DSA} that $L$ stands for the input's max load. Thus if Corollary 17 were a function, its input would be a set of jobs, and its output would be a set of valid offsets with which to annotate the input. Moreover, its body would comprise (i) a computation of $\epsilon$ and (ii) an invocation of Theorem 16.

Though Corollary 17 proved inappropriate as an entry point, it was useful in the sense of fixing our attention to Theorem 16. To that FU we now turn~\footnote{The numbering of FUs in the original BA publication carries an implicit indication of strength, i.e., width of applicability and/or degree of approximation. Lemma 1 operates on unit-size jobs that are all live at the same time. Theorem 2 treats unit-size jobs with arbitrary lifespans, thus removing the simultaneous liveness constraint and widening its applicability. Theorem 16 deals with arbitrary input sets and guarantees solutions with makespan at most $(1+c\epsilon)L~+~O(h_{max}/\epsilon^6)$ for some constant $c$ and some real $\epsilon$. The strongest algorithm in the paper is featured in Theorem 19, which nevertheless cannot be implemented as a computer program (see the Appendix for an elaboration).}. As regards its proof, we omit mathematical arguments in between computational steps. We make omissions explicit via the symbol ``[...]''.

\begin{t16}
Let $\epsilon \in (0,1]$. There exist a constant $c$ and a polynomial-time algorithm that takes $\epsilon$ and an arbitrary set $X$ of jobs as input and produces a feasible solution to DYNAMIC STORAGE ALLOCATION on $X$ with makespan at most $(1+c\epsilon)L~+~O(h_{max}/\epsilon^6)$.
\end{t16}

\begin{proof}    
    $[...]$ We are going to apply Corollary 15 repeatedly, boxing the smallest jobs so as to increase the minimum job height $h_{min}$ until it gets close enough to the maximum job height $h_{max}$ that we can finish with a last application of Corollary 15.

    $[...]$ Let $r$ denote the ration $h_{max} / h_{min}$. Assume first that $(log_2r)^2 \geq 1/\epsilon$, and set $\mu=\epsilon/{(log_2r)^2}$ and $H=\lceil \mu^5h_{max}/(log_2r)^2\rceil$. Consider the partition $X=X_s~\cup~X_l$, where $X_s$ denotes the jobs of height at most $\mu H$ and $X_l=X~\setminus~X_s$. Now apply Corollary 15 to $X_s$ with box-height parameter $H$ and error parameter $\mu$. This yields a set $B_s$ of boxes of height $H$ into which the jobs of $X_s$ fit such that $[...]$.

    Now consider $B_s$ as a set of jobs and the revised problem on $X'=B_s~\cup~X_l$. $[...]$ Iterate the above boxing of small jobs, each time using new error parameter $\mu'=\epsilon/(log_2r')^2$ until it yields a problem $X^*$ with minimum job height $h_{min}^*$ for which the ratio $r^*=h_{max}/h_{min}^*$ is such that $(log_2r^*)^2 < 1/\epsilon$. $[...]$

    Now apply Corollary 15 to all of $X^*$ with box-height parameter $H=h_{max}/\epsilon$ $[...]$ and error parameter $\epsilon$; this is the ``last application'' of Corollary 15 to which we alluded earlier. $[...]$
\end{proof}

It must now be obvious that Theorem 16 is the crux of BA, i.e., its ``main'' function. It accepts an arbitrary set of jobs and a real number, and produces the corresponding DSA solution. The following remarks apply:

\begin{itemize}
    \item the execution of Theorem 16 is governed by $\epsilon$, $h_{min}$ and $h_{max}$. Everything else is a function of these three quantities.
    
    \item the actual output of Theorem 16 is not a complete DSA solution. As we can see, the proof is built around repeated applications of Corollary 15, and terminates with such an application. According to the proof's own phrasing, however, Corollary 15 produces \textit{boxes}; not offsets.
    
    \item the loop that is executed while $(log_2r)^2 \geq 1/\epsilon$ demands that $h_{min} \leq \mu H$, else $X_s$ turns out empty. Then $B_s$ is empty as well, the ratio $r$ remains unchanged, and the loop never ends.
\end{itemize}

The first remark is self-explanatory. As regards the second remark, its validity does not harm the purpose of BA's authors. Their argument rather exploits the fact that Corollary 15 produces same-height boxes. Recall from Section \ref{sec:elem} that IGC applied on identical sizes yields zero fragmentation, i.e., the solution's makespan equals the input's max load. In the parts of the proof that we have omitted for brevity, the authors bound the max load of Corollary 15's output, thus bounding the makespan of the boxes' contents as a result. From the perspective of a programmer who wants to actually solve DSA, implementing Theorem 16 is insufficient. Hence the first paragraph of the present subsection.

The final remark is in fact the opening of the rabbit hole which led us to discovering BA's latent invariants.

\subsubsection*{Interlude: Programming as Archaeology}
Allow us to clarify our stance before proceeding. From the outset of our efforts to this day, we have put our ultimate trust on BA's superiority. We view its FUs as priceless ancient artifacts buried in the sands of abstract thought, and our work as that of an archaeologist who must unearth those artifacts in the most intact form possible. This act of excavation, this transition from theory to practice, from the abstract to the executable, unavoidably entails points of necessary intervention. Our unshakeable trust on BA dictates (i) minimizing the number and degree of said interventions, as well as (ii) being certain about their soundness. It is these two implications that the following Section serves. A formal treatment of our findings is beyond both our powers and intentions.

\subsection{Latent Invariants}
\label{sec:latent}
By this Section's title we are referring to the following non-trivial conclusions:

\begin{enumerate}
    \item the real-valued $\epsilon$ of Theorem 16 has an input-dependent range of ``legal'' values which can be greater than 1.

    \item the real-valued $\mu$ of Theorem 16 has a universal upper bound equal to ${\sqrt{5} - 1}\over2$.

    \item it is necessary that every input satisfies the inequality $h_{max} \geq \lceil 2216.53\cdot h_{min}\rceil$.
\end{enumerate}

We shall show that all three invariants can be derived from BA's original text without any additional moves. Let us start with some definitions from the proof of Theorem 16, particularly that branch of execution where $(log_2r)^2 \geq 1/\epsilon$:

\begin{equation}
    \label{eq:r}
    r = {h_{max}\over h_{min}}
\end{equation}
\begin{equation}
    \label{eq:mu}
    \mu={\epsilon \over (log_2r)^2}
\end{equation}
\begin{equation}
    \label{eq:H}
    H=\lceil \mu^5h_{max}/(log_2r)^2\rceil
\end{equation}

As we have already remarked, in order for that branch to avoid looping forever, it should hold that $h_{min} \leq \mu H$. Let us unwrap this expression:

\begin{align*}
    h_{min} &\leq \mu H \xRightarrow{\eqref{eq:H}} \\
    h_{min} &\leq \mu \lceil \mu^5h_{max}/(log_2r)^2\rceil \xRightarrow{\mu~>~0} \\
    {h_{min}\over\mu} &\leq \lceil \mu^5h_{max}/(log_2r)^2\rceil \xRightarrow[]{}\\
    {h_{min}\over\mu} - 1 &< \mu^5h_{max}/(log_2r)^2
\end{align*}

To the last equation, we can without loss of generality tighten its left hand:

\begin{align*}
    {h_{min}\over\mu} - 1 &< \mu^5h_{max}/(log_2r)^2 \xRightarrow{}\\
    {h_{min}\over\mu} &\leq \mu^5h_{max}/(log_2r)^2 \xRightarrow[log_2r > 0]{\eqref{eq:r}}\\
    {(log_2r)^2\over r} &\leq \mu ^6 \xRightarrow{\eqref{eq:mu}}\\
    {(log_2r)^2\over r} &\leq {\epsilon^6\over (log_2r)^{12}} \xRightarrow{}
\end{align*}
\begin{equation}
    \label{eq:esmall}
    \epsilon \geq \sqrt[6]{(log_2r)^{14}\over r}
\end{equation}

We have arrived at a condition for $\epsilon$ which, given the fact that Theorem 16 operates on \textit{arbitrary} sets of jobs, i.e., for any $r$, does by no means guarantee that $\epsilon \in (0,1]$. Let us move forward. Recall that we are undergoing this investigation in order to arrive at conditions which guarantee that the algorithm described in the proof of Theorem 16 runs ``as it should''. Also recall that we are for now focusing on the top branch of said proof, namely that one where $(log_2r)^2 \geq 1/\epsilon$. There, Corollary 15 is called on $X_s$ with box-height parameter $H$ and error parameter $\mu$. Here's Corollary 15:

\begin{c15}
Let $H$ be a positive integer box-height parameter and $\epsilon >0$ be a sufficiently small error parameter. Given a set $Z$ of jobs, each of height between $h_{min}$ and $\epsilon H$, there exist a set $B$ of boxes, each of height $H$, and a boxing of $Z$ into $B$ such that for all x-coordinates $t$,
\begin{equation*}
    L_B(t) \leq (1+9\epsilon)L_Z(t) + O({{H(log_2(H/h_{min}))^2}\over \epsilon^4})
\end{equation*}
\end{c15}
\begin{proof}
    We construct such a boxing. First, round the job heights: each height $h$ is rounded up to $\lfloor (1+\epsilon)^i\rfloor$, where $i$ is defined by $(1+\epsilon)^{i-1}<h\leq (1+\epsilon)^i$. Let $Y$ denote the resulting set of rounded jobs.

    Now, partition the jobs according to their heights. For each rounded height $h$, let $Y_h$ denote the set of jobs of height $h$. Divide the heights of all jobs in $Y_h$ by $h$; apply Theorem 2 with box-height parameter $\lfloor H/h\rfloor$; and then multiply all box heights by $h$ to get a set $B_h$ of boxes of height at most $H$. The output is a set $B=\bigcup_h B_h$ of boxes, which we can assume are all of height $H$. $[...]$
\end{proof}

A non-obvious yet key detail is that we must not call Theorem 2 with a box-height parameter equal to zero (since zero-height boxes do not make sense). We see from the proof that that box-height parameter is determined by the size classes to which the input jobs have been rounded up. We know that there exists a $i_{max}$ for which the largest jobs in $Z$ are rounded to $h_m=\lfloor(1+\epsilon)^{i_{max}}\rfloor$. It suffices to ensure $\lfloor H/h_m\rfloor\geq1$:

\begin{align*}
    \lfloor H/h_m\rfloor &\geq 1 \Rightarrow\\
    H/h_m &\geq 1 \Rightarrow\\
    h_m &\leq H \Rightarrow\\
    \lfloor(1+\epsilon)^{i_{max}}\rfloor &\leq H \Rightarrow\\
    (1+\epsilon)^{i_{max}} &< H + 1 \xRightarrow{\epsilon~=~\mu}
\end{align*}
\begin{equation}
    \label{eq:mumin}
    (1+\mu)^{i_{max}} < H + 1
\end{equation}

Due to the fact that Corollary 15 is called on $X_s$ ($Z=X_s$), we know that for all sizes $h$ in $Z$:

\begin{equation}
    \label{eq:c15h}
    h_{min} \leq h \leq \lfloor \mu H\rfloor
\end{equation}

We can thus expand Inequality \ref{eq:mumin} with another branch on its left side, since $\lfloor \mu H\rfloor \leq h_m$:

\begin{align*}
    \lfloor \mu H\rfloor &\leq (1+\mu)^{i_{max}} < H + 1 \Rightarrow\\
    \lfloor \mu H\rfloor &< H + 1 \Rightarrow\\
    \mu H &< H + 1 \Rightarrow\\
    H(1-\mu) &> -1
\end{align*}

The above is always true as long as $1-\mu \geq 0 \Rightarrow \mu \leq 1$. A rather sensible requirement given the fact that, overall, Corollary 15 boxes jobs of height up to $\mu H$ into $H$-sized boxes.

Before examining Theorem 2, let us backtrack to consider the second execution path of Theorem 16, that where $(log_2r)^2 < 1/\epsilon$. BA's authors suggest to invoke Corollary 15 one last time, with box-height parameter $H=h_{max}/\epsilon$ and error parameter $\epsilon$. Our analysis, however, forces us to reject this course of action. We have already shown that (i) $\epsilon$ may end up greater than 1 and (ii) Corollary 15 demands an error parameter that is \textit{at most} 1. An alternative is necessary.

The repeated applications of Corollary 15 during the top branch of Theorem 16 increase the minimum job height to $h_{min}^*$. We thus know that $r^*=h_{max}/h_{min}^*$ is smaller than all the previous values of $r$. As a result, $\mu^*=\epsilon/(log_2r^*)^2$ is the \textit{maximum} value for $\mu$. What if we used $\mu^*$ in the place of $\epsilon$ for the last invocation of Corollary 15? Similarly with before, we would have:

\begin{equation}
    \label{eq:mustar}
    (1+\mu^*)^{i_{max}-1}<h_{max}\leq (1+\mu^*)^{i_{max}}
\end{equation}

Demanding that the largest size class does not yield a zero box-height parameter for Theorem 2 leads us to:

\begin{align*}
    \lfloor(1+\mu^*)^{i_{max}}\rfloor &\leq H \Rightarrow\\
    (1+\mu^*)^{i_{max}} &< H + 1 \xRightarrow{H=h_{max}/\mu^*}\\
    (1+\mu^*)^{i_{max}} &< {h_{max}\over \mu^*} + 1
\end{align*}

To simplify our algebra, we can once again without loss of generality prune the last inequality to $(1+\mu^*)^{i_{max}} \leq {h_{max}\over \mu^*}$. Dividing all members of Inequality \eqref{eq:mustar} with $\mu^*$ and keeping the left side, we have ${(1+\mu^*)^{i_{max}-1}\over\mu^*}<{h_{max}\over\mu^*}$. We must now decide about the relation between $(1+\mu^*)^{i_{max}}$ and ${(1+\mu^*)^{i_{max}-1}\over\mu^*}$. Nothing obstructs us from declaring the below:

\begin{align*}
    (1+\mu^*)^{i_{max}} &\leq {(1+\mu^*)^{i_{max}-1}\over\mu^*} \Rightarrow\\
    (1+\mu^*)\mu^* &\leq 1 \Rightarrow\\
    \mu^{*2} + \mu^* - 1 &\leq 0
\end{align*}

The corresponding equation has roots $\mu^*_{1,2}={{-1 \pm \sqrt{5}}\over 2}$. Since $\mu^*$ is by definition positive, the only way for the inequality to be less or equal than zero is:

\begin{equation}
    \label{eq:mufin}
    \mu^* \leq {{\sqrt{5}-1}\over 2} \simeq 0.618033...
\end{equation}

This is a very convenient result. First of all, it abides to our requirement with respect to the error parameter given to Corollary 15. In other words, we \textit{can} use $\mu^*$ instead of $\epsilon$ for the last invocation of Corollary 15, as long as Inequality \ref{eq:mufin} holds. Secondly, it is independent from the input. The only problem is, $\mu^*$ is a quantity ``from the future'': BA has to execute properly and reach the low branch of Theorem 16 before $r^*$---and thus $\mu^*$---becomes available. In contrast, we want to control BA's execution via configuring quantities that are available from the outset, like $\epsilon$ and $r$. Thankfully, $\mu^*$ is a function of $\epsilon$. Having decided to use $\mu^*$ for the last Corollary 15 invocation, and knowing the necessary condition for this to work (Inequality \ref{eq:mufin}), we can impose it to $\epsilon$ in the here and now:

\begin{align*}
    \mu^* &\leq {{\sqrt{5}-1}\over 2} \xRightarrow{\mu^*={\epsilon\over (log_2r^*)^2}}\\
    \epsilon &\leq {{\sqrt{5}-1}\over 2}(log_2r^*)^2 \xRightarrow[\eqref{eq:esmall}]{r^*~<~r}
\end{align*}
\begin{equation}
    \label{eq:efin}
    \sqrt[6]{(log_2r)^{14}\over r} \leq \epsilon \leq {{\sqrt{5}-1}\over 2}(log_2r)^2
\end{equation}

There is, however, no reason to believe that Inequality \ref{eq:efin} will be valid for \textit{all} possible inputs. In order to be certain we must make one last demand:

\begin{align*}
     \sqrt[6]{(log_2r)^{14}\over r} &< {{\sqrt{5}-1}\over 2}(log_2r)^2 \Rightarrow\\
     {(log_2r)^{14}\over r} &< ({{\sqrt{5}-1}\over 2})^6 \cdot (log_2r)^{12} \Rightarrow
\end{align*}
\begin{equation}
    \label{eq:rfin}
    {(log_2r)^2\over r} < ({{\sqrt{5}-1}\over 2})^6
\end{equation}

According to \href{https://www.wolframalpha.com/input?i=log2%28x%29%5E2%2Fx+%3C+%28%28sqrt%285%29-1%29%2F2%29%5E6}{WolframAlpha}, an approximate solution for Inequality \ref{eq:rfin} is $r>2216.53$. This concludes our design. Inequalities \ref{eq:efin}, \ref{eq:mufin} and \ref{eq:rfin} correspond to each of the three invariants listed in the beginning of this Section. Incorporating them to our source code has allowed \texttt{idealloc} to treat a wide variety of inputs without any unexpected behavior.

\subsection{Critical Point Injection}
\label{sec:cpi}
The latent invariants of the preceding Section do the ``heavy lifting'' of ensuring that BA works as it should. Our tour, however, is not over. There is one last intervention that we needed to make. It is time to visit Theorem 2:

\begin{t2}
Given a set $Z$ of jobs, each of height 1, an integer box-height parameter $H$, and a sufficiently small positive $\epsilon$, there exist a set $B$ of boxes, each of height $H$, and a boxing of $Z$ into $B$ such that for all x-coordinates $t$,

\begin{equation*}
    L_B(t) \leq (1+4\epsilon)L_Z(t) + O({Hlog_2H\over \epsilon^2}log_2{1\over\epsilon})
\end{equation*}
\end{t2}

\begin{proof}
    We are going to apply Lemma 1 many times, boxing the unresolved jobs into additional boxes as we go along. Our general goal is to keep the wasted load (free space) in those additional boxes small at any x-coordinate.

    We use the following recursive method. Given are

    \begin{itemize}
        \item A set $X$ of jobs and an open \textit{bounding interval} $I$, such that $\forall j~\in X,~I_j\subseteq I$.

        \item A nonempty finite set of \textit{critical x-coordinates} $T=\{infI=t_o < t_1 <~...~<t_q < t_{q+1}=supI\}\subseteq I~\cup\{infI,~supI\}$.

        \item A set $F$ of \textit{free spaces}. Each free space is an open sub-interval of $I$ of height 1 \textit{having endpoints in $T$}. Any free space $f\in F$ is called \textit{spanning} if $f=I$ and \textit{non-spanning} otherwise.
    \end{itemize}

    Initially, $X=Z$, $I=(0,~1)$, $T=\{0,~t,~1\}$ for some arbitrary $t$ at which some job from $Z$ is live, and $F=\emptyset$. Recall that $I_j=(x_j,~y_j)$ denotes the interval of job $j$. With the help of $T$, define partition

    \begin{equation*}X=(R_1~\cup~R_2~\cup~...~\cup~R_q)~\cup~(X_0~\cup~X_1~\cup~...~\cup~X_q)
    \end{equation*}

    as follows. First, define $X_i=\{j\in X: I_j \subseteq(t_i,~t_{i+1})\}$ for $0\leq j \leq q$.

    Then define the $R_i$'s recursively. Define $X'=X~\setminus~(X_0~\cup~X_1~\cup~...~\cup~X_q)$. Note that $q\geq 1$. Define $R_{\lceil q/2\rceil}=\{j\in X':~t_{\lceil q/2\rceil} \in I_j\}$. Define $P$ to be the set of remaining jobs $j$ of $X'$ with $y_j < t_{\lceil q/2\rceil}$, and define $Q$ to be the set of remaining jobs $j$ of $X'$ with $t_{\lceil q/2\rceil} < x_j$. If $P\ne\emptyset$, recursively partition $P$ using $\{t_1,~t_2,~...~,t_{\lceil q/2\rceil-1}\}$. Afterward, if $Q\ne \emptyset$, recursively partition $Q$ using $\{t_{\lceil q/2\rceil+1},~t_{\lceil q/2\rceil+2},~...,~t_q\}$.

    Now to each $X_i$ associate a set $F_i$ of intervals (free spaces), initially empty. As sections of free spaces in $F$ are used to box jobs in the $R_i$'s, the unused fragments will be deposited into the appropriate $F_i$'s for use deeper in the recursion (to box jobs in the $X_i$'s).

    To box the jobs in the $R_i$'s, first apply Lemma 1 to each $R_i$, $1\leq i \leq q$, in any order; note that all jobs in $R_i$ are live at $t_i$. For each $i$, this boxes all the jobs of $R_i$ except for at most $2H\lceil1/\epsilon^2\rceil$ unresolved jobs. Now consider the set $U$ of all the unresolved jobs from all the $R_i$'s. Derive an optimal packing of $U$ using interval graph coloring (Recall that all jobs are of height one). This packing has makespan $L_U$.

    Let $s(F)$ denote the subset of spanning free spaces of $F$. If $|s(F)| < L_U$, create $\lceil (L_U-|s(F)|)/H\rceil$ boxes of height $H$ and horizontal extent $I$. This yields $H\lceil (L_U-|s(F)|)/H\rceil$ new spanning free spaces; add them to $F$. Now there are at least as many spanning free spaces in $F$ as rows of the packing of $U$.

    For each $1\leq j\leq L_U$, remove one spanning free space from $F$, and use it to place all the jobs in row $j$of the packing. This creates \textit{gaps}, or unused portions, in the original free space, each of the form $[\alpha,~\beta]$ where for some $i$, $j$: $t_i < \alpha < t_{i+1}$ and $t_j < \beta < t_{j+1}$; recall that $t_0=infI$ and $t_{q+1}=supI$. For each such $[\alpha,~\beta]$, if $i\ne j$ then split $[\alpha,~\beta]$ into $(\alpha,~t_{i+1})$, $(t_{i+1},~t_{i+2})$, $...$, $(t_{j-1},~t_j)$, $(t_j,~\beta)$; and add $(\alpha,~t_{i+1})$ to $F_i$, $(t_{i+1},~t_{i+2})$ to $F_{i+1}$, $...$, $(t_{j-1},~t_j)$ to $F_{j-1}$, and $(t_j,~\beta)$ to $F_j$. Otherwise ($i=j$), simply deposit $(\alpha,~\beta)$ into $F_i$. This \textit{fragments} the gaps.

    Now all the jobs in all the $R_i$'s are boxed. Consider the unused free spaces in $F$, if any. Each is of the form $(t_i,~t_j)$ for some $i\ne j$. Split each such $(t_i,~t_j)$ into $(t_i,~t_{i+1})$, $(t_{i+1},~t_{i+2})$, $...$, $(t_{j-1},~t_j)$. Add $(t_i,~t_{i+1})$ to $F_i$, $(t_{i+1},~t_{i+2})$ to $F_{i+1}$, $...$, and $(t_{j-1},~t_j)$ to $F_{j-1}$. This \textit{passes down} the remaining unused free spaces to the sub-problems.

    In parallel for each $\ell=0,~1,~2,~...,~q$, if $X_\ell\ne\emptyset$, recursively apply the construction with new $X\leftarrow X_\ell$, new free space set $F\leftarrow F_\ell$, new bounding interval $I\leftarrow (t_\ell,t_{\ell+1})$ and new criticall x-coordinate set $T\leftarrow \{$endpoints of elements of $F_\ell$$\}~\cup~\{t_\ell,~t_{\ell+1}\}$. $[...]$
\end{proof}

\begin{table*}
    \caption{Findings and remedies applied to BA's FUs.}
    \label{tab:ba_interventions}
    \centering
    \begin{tabular}{||p{0.11\textwidth}|p{0.45\textwidth}|p{0.4\textwidth}||}
        \hline
         \textbf{BA FU} & \textbf{Finding} & \textbf{Remedy} \\ \hline
         \multirow{2}{*}{Corollary 17} & Does not comply with latent invariants of Theorem 16 and Corollary 15. & \multirow{2}{=}{Input preprocessing, $\epsilon$-calibration (Section \ref{sec:prelude}).} \\ \hline
         \multirow{3}{*}{Theorem 16} & 1. Last Corollary 15 invocation uses $\epsilon$ (unsafe).\newline\newline2. Yields boxes instead of offsets. & \multirow{3}{=}{1. Use $\mu^*$ instead (Section \ref{sec:latent}).\newline\newline2. Unbox and place (Section \ref{sec:unbox}).} \\ \hline
         Theorem 2 & $R$ can be empty. & Critical point injection (Section \ref{sec:cpi}). \\ \hline
         Corollary 15 & \multirow{2}{*}{As is, as long as the above remedies are applied.} & \multirow{2}{*}{N/A} \\ \cline{1-1}
         Lemma 1 & & \\ \hline
    \end{tabular}
\end{table*}

By now our initial point that BA is not simply an ``algorithm'' must be obvious. We discourage the reader from devoting excess effort to grasping every last word of Theorem 2 (as we shall show in Section \ref{sec:design}, some parts of it are redundant). For the time being, it suffices to pay attention to the fact that in order for the boxing procedure to advance, there must exist at least one critical x-coordinate in $T$ at which at least one job in $Z$ is live. In other words, there must exist at least one $R_i$. At each recursion level, it is only jobs in $R_i$'s that are being boxed, some via Lemma 1, and others via IGC. This need is made explicit at the start of the proof, where attention is drawn to ``some arbitrary $t$ at which some job from $Z$ is live''. In our experience, however, it is possible deeper in the recursion for critical point sets $T$ to appear carrying no such $t$. In those cases, we append one more (appropriate) time point to $T$.

The only remaining FU in BA's chain is Lemma 1. To keep the main body of our paper as short as possible, and due to the fact that Lemma 1 works ``out of the box'', we have moved its definition to the Appendix.

To summarize, the entire Section \ref{sec:coreba} demonstrates our approach as regards \texttt{idealloc}'s core component, namely the boxing algorithm by Buchsbaum et al.~\cite{buchsbaum}. We have gone through the algorithm's parts and limitations, and have either presented, or hinted toward, ways to overcome said limitations. The main takeaways are listed in Table \ref{tab:ba_interventions}.

\section{Unboxing and Final Placement}
\label{sec:unbox}
We have already mentioned that BA does not produce offsets, as would normally be the case if one wanted to solve DSA. Instead Theorem 16 returns a set of equal-height, Matryoshka doll-like boxes. The problem addressed by the present Section can be stated as: \textit{how can the outer Matryoshkas' IGC-derived offsets be diffused all throughout the boxing's hierarchy until the original buffers are found and accordingly placed?}

The process is sketched in Figure \ref{algo:placement}. We will be using ``buffers'' to refer to original buffers and ``boxes'' for the Matryoshkas. Like boxing, this is a recursive procedure. Two questions are driving decisions at each level of recursion:

\begin{itemize}
    \item \textbf{Line 3:} do input elements share the same size?
    \item \textbf{Line 5:} are input elements non-overlapping?
\end{itemize}

\begin{wrapfigure}[39]{r}{0.53\textwidth}
    \begin{minipage}{0.53\textwidth}
        \begin{algorithm}[H]
            \SetKwProg{Fn}{Function}{}{end}
            \SetKwInOut{Input}{input}
            \SetKwInOut{Output}{output}
            \SetKwFunction{Place}{UnboxAll}
            \SetKwData{Water}{w}
            \SetKwFunction{JB}{JustBoxes}
            \SetKwFunction{JBU}{JustBuffers}
            \SetKwFunction{Over}{Overlap}
            \SetKwFunction{IG}{IGC}
            \SetKwFunction{UB}{Unbox}
            \SetKwFunction{SS}{SameSize}
            \SetKwFunction{MA}{MaxAddr}
            \SetKwFunction{Init}{Init}
            \SetKwFunction{Merge}{MergeOffsets}
            \SetKwFunction{BA}{BumpByArea}   
            \SetKwFunction{PA}{PlaceAt}
            \SetKwFunction{Part}{PartitionBySize}
            \SetKwFunction{V}{OffsetsAreValid}
            \SetKwData{R}{row}
            \SetKwData{P}{placed}
            \SetKwData{B}{job}
            \SetKwData{SC}{size}
            \SetKwData{J}{jobs}
            \SetKwFunction{OS}{PlaceSameSizes}
            \Fn{\Place{$J$, \Water}}{
            \tcp{A set of BA-derived jobs and a} \tcp{starting offset.}
            \Input{$J = \{j | j = (h, t_s, t_e)\}$, \Water}
            \tcp{A valid offset-job mapping.}
            \Output{$O = \{o| o \in \mathbb{N} : \V{J, O}\}$}
            \BlankLine
            $O \leftarrow$ \Init{}\;
            \uIf{\SS{$J$}}{
                \KwRet{\OS{$J$,\Water}}\;
            }
            \uElseIf{not \Over{$J$}}{
                \For{\B in $J$}{
                    \P $\leftarrow$ \Place{\UB{$J$}, \Water}\;
                    $O \leftarrow$ \Merge{$O$, \P}\;
                }
            }
            \Else{
                \For{\J in \Part{$J$}}{
                    \P $\leftarrow$ \OS{\J, \Water}\;
                    \Water $\leftarrow$ \MA{\P}\;
                    $O \leftarrow$ \Merge{$O$, \P}\;
                }
            }
            \KwRet{$O$};
            }
            \BlankLine
        
            \Fn{\OS{$J$,\Water}}{
            \Input{$J = \{j | j = (h, t_s, t_e)\}$, \Water}
            \Output{$O = \{o| o \in \mathbb{N} : \V{J, O}\}$}
            \BlankLine
            $O \leftarrow$ \Init{}\;
            \tcp{All jobs have the same size.}
            \For{\R in \IG{$J$}}{
                \P $\leftarrow$ \Place{\R,\Water}\;
                \Water $\leftarrow$ \MA{\P}\;
                $O \leftarrow$ \Merge{$O$, \P}\;
            }
            \KwRet{$O$}\;
            }
        \end{algorithm}               
    \end{minipage}
    \caption{Unboxing pseudocode.}
    \label{algo:placement}
\end{wrapfigure}

Apart from buffers/boxes, a watermark is also given as input---initialized at zero before the first ever call. It signifies the starting offset from which placement should commence. The watermark is updated and inherited by deeper recursion levels. Hence we ensure that the contents of each box end up placed within their container's boundaries.

Let us now visit all possible answers to the above questions. If jobs share the same size, we exploit the fact that DSA for uniform sizes is optimally solved with IGC. The role of \texttt{PlaceSameSizes} is to traverse all IGC-produced rows, place the contents of each at the current watermark, and bump the watermark at the row's tip. If the jobs don't overlap in time, the decision is trivial. We unbox each input element and recursively call the procedure with the same watermark (lines 7, 8).

Finally, if none of the above conditions hold, we partition the jobs by size and place each subset independently (lines 12-14). But we are not done yet! Due to the repeated round-ups of box sizes in Corollary 15, as well as the recursive nature of Theorem 16, the offsets produced by the above procedure are \textit{sparse}. So we view all work up to this point, i.e., BA-derived boxing and the offsets produced by unboxing, as an intricate sorting step according to the terminology of Section \ref{sec:heur}. To finalize the output, we ``squeeze'' the buffers via first-fit placement, traversing them in increasing offset.

\section{Design and Implementation}
\label{sec:design}

\begin{figure}[t!]
    \centering
    \includegraphics[width=0.8\textwidth]{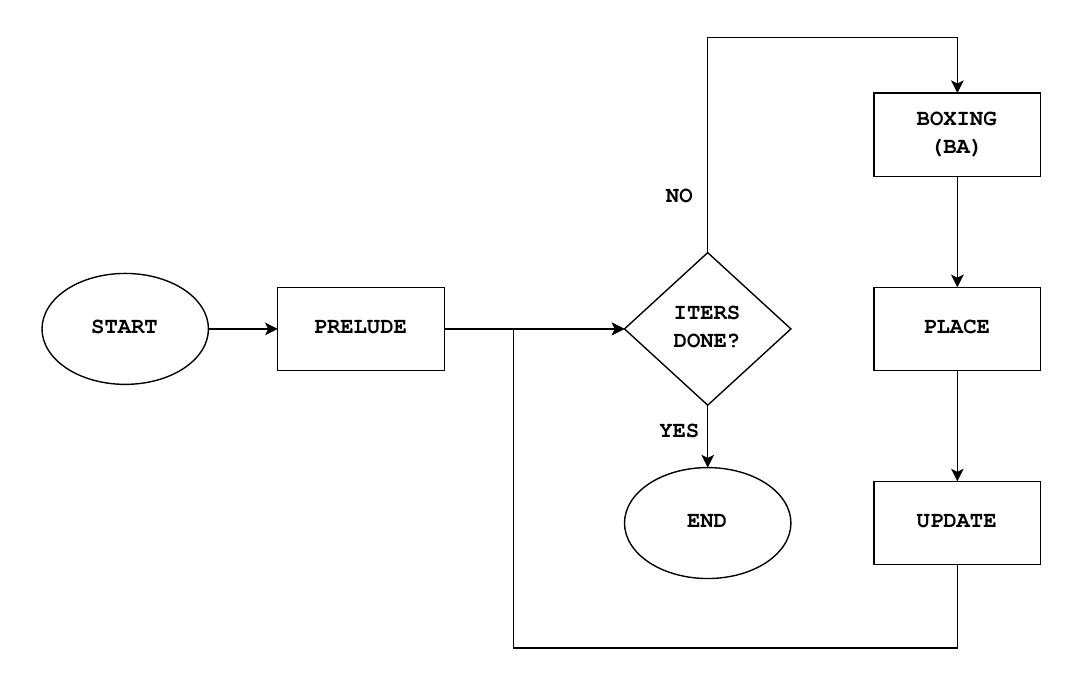}
    \caption{\texttt{idealloc} flow diagram.}
    \label{fig:idealook}
\end{figure}

Figure \ref{fig:idealook} gives an overview of \texttt{idealloc}. Its design is owed (i) to our goal of robust and high performance, and (ii) to the inherent \textit{stochasticity} of BA, due to the critical points of Theorem 2 (see Section \ref{sec:cpi}). Though we have more to say on this later, keep in mind that even the simplest of operations, namely sorting by size, is stochastic: how should one break ties between equal sizes? Enforced determinism, i.e., using some unique ID for such occasions, may help with data visualization but harms best-case fragmentation. One does not tame randomness by putting it under the rug.

\subsection{Interface}
\label{sec:interface}

\texttt{idealloc} accepts the following parameters:

\begin{itemize}
    \item \textbf{original input:} a collection of jobs to place.
    \item \textbf{worst-case fragmentation:} an upper bound for the quality of the output. If that amount or less fragmentation is achieved at any point, the execution terminates early.
    \item \textbf{start address:} base location $S$ to which all offsets refer. The address of a buffer with offset $O$ is $S+O$.
    \item \textbf{iterations:} an upper bound for the total number of times the box-and-place kernel is allowed to run. If exhausted and worst-case fragmentation is not yet beaten, the next-best result is returned.
\end{itemize}

Note that, as regards fragmentation, in our opinion the only optimal value is zero. But we have included the respective parameter in response to allocators like \texttt{minimalloc}~\cite{minimalloc} and the one featured in Apache's TVM compiler, who include a ``maximum makespan'' parameter to their interfaces. We find it erroneous to decouple worst-case storage from the input, since it is the input itself, and specifically its max load, which bounds makespan (from below, not from above). Certain maximum makespans may not be achievable for certain inputs.

\subsection{Input Representation}
\label{sec:job}
The fundamental data structure of \texttt{idealloc} is the \texttt{Job}. Its fields are:

\begin{itemize}
    \item \textbf{allocated size:} self-explanatory.
    \item \textbf{(start, end):} the respective allocation and deallocation times. In line with DSA theory, we adopt \textit{exclusive lifetime semantics} in \texttt{idealloc}. This means that a job is \textbf{not} live at neither its start, nor its end. Numerous bugs have crunched our nighttime due to ours not being strict enough about lifetime semantics.
    \item \textbf{alignment:} if any, the final address of the buffer is guaranteed to be a multiple of this value.
    \item \textbf{requested size:} owed to the beginnings of \texttt{idealloc} being in studying \texttt{malloc} traces, kept because someone else may decide to do so in the future. By knowing the difference between requested and allocated size one can measure \textit{internal} fragmentation, out of scope for this paper.
    \item \textbf{contents:} a job may be a box spawned by BA, holding other jobs inside. Both such boxes \textit{and} the original buffers of the input are represented with the same struct.
    \item \textbf{id:} self-explanatory.
\end{itemize}

Some further remarks on how we handle the input. First of all, there is a list of security checks that must be conducted \textit{before} \texttt{idealloc} is invoked. Zero-valued sizes are not allowed. Start- equal or greater than end-times are not allowed. Zero-valued alignment (different than \textit{no} alignment) is not allowed. Non-empty contents are not allowed. Last but not least, we do not allow allocated sizes to be smaller than requested sizes.

\subsection{Event Traversal}
\label{sec:evts}
A common situation in \texttt{idealloc} is that of computations operating on subsets of buffers. In our experience, avoiding quadratic complexity in such cases is crucial to the allocator's execution time and scalability. Take the max load $L$ of Section \ref{sec:DSA} as an example. Recall that $L$ amounts to the maximum amount of memory that is concurrently live at any time. A naive quadratic solution is to traverse all allocation and deallocation times of all buffers, and for each one traverse the buffers themselves, and aggregate the sizes of those that are live. Luckily there is a better approach.

Imagine a priority queue consisting of \textit{events}: each event carries (i) a timestamp, (ii) a type, i.e., whether it marks the allocation or deallocation of a job, and (iii) a reference to the job itself. Earlier events have precedence over later ones, and deallocations have precedence over allocations. The max load $L$ of $N$ buffers can be computed by consuming this priority queue once, thus by processing $2N$ events. We make heavy use of event traversal across \texttt{idealloc} and consider it a fundamental operation. Its underlying principle is that \textit{no change of any kind occurs between consecutive events}.

\subsection{Working with Different Lifetime Semantics}
\label{sec:sem}
Fellow allocators and/or benchmarks ascribe different interpretations to buffers' intervals. For instance, XLA's best-fit heap simulator views jobs as live at the endpoints as well as the in-between. \texttt{minimalloc} is start-inclusive end-exclusive. \texttt{idealloc} adopts exclusive semantics for its internal operation.

Suppose the very real scenario of needing to conduct the experiments accompanying this paper. Given the aforementioned variety of semantics in the SOTA, one needs to be certain that they are comparing apples to apples. In other words, allocators with different semantics must agree, regarding the buffers described by a specific input benchmark, on which pairs of buffers do or do not overlap. A necessary but not sufficient condition when pursuing such an agreement is that the reported max load of the \textit{same} dataset expressed in exclusive semantics be equal to the one reported when using any other semantics. We make active use of this check in our measurement scripts.

Assume we are in possession of a benchmarks suite employing start-inclusive, death-exclusive semantics. We will be referring to this interpretation as \texttt{InEx} from now on, and will be using \texttt{In} and \texttt{Ex} for start-inclusive-end-inclusive and start-exclusive-end-exclusive semantics respectively. Assume, further, that we want to evaluate on this suite three allocators: the first uses \texttt{InEx}, the second \texttt{In}, and the last one \texttt{Ex}. Last but not least, assume that the task of reading a DSA solution, validating its feasibility, and reporting statistics of interest such as its max load and makespan, is carried out by an analyzer program also using \texttt{Ex} semantics. This description largely resembles our real experiments setup.

The missing component is an \textit{adapter}, its input being (i) a DSA solution file, (ii) the semantics of that file and (iii) the semantics to which the file's contents must be transformed. By making use of this adapter, we can for example start from an \texttt{InEx} dataset, feed it to the \texttt{In}-allocator, and then pass its output to the \texttt{Ex}-analyzer. Regardless from the point of departure, the analyzer must always report the same max load and the same number of conflicts (i.e., distinct pairs of overlapping buffers) for the same benchmark. The \texttt{idealloc} source code includes such an adapter. Its operating principles are:

\begin{itemize}
    \item \texttt{In} $\longleftrightarrow$ \texttt{InEx}: add or subtract one from the buffer's de-allocation time, depending on the direction of the arrow
    \item \texttt{InEx} $\longleftrightarrow$ \texttt{Ex}: the two types are \textit{equivalent}. The condition for conflict with a buffer allocated at $a$ and de-allocated at $b$ is in both cases $\neg(x\leq a~\lor~y\geq b)$, where $x,~y$ stand for the respective endpoints of some other buffer
\end{itemize}

\subsection{Bootstrapping and Early Stopping}
\label{sec:bootstr}
Due to its stochastic nature, the quality of solutions that \texttt{idealloc} may yield at each iteration exhibits great variety. In order to waste as little time as possible on sub-optimal solutions, we use a simple bootstrapping scheme: we keep a record of the smallest makespan achieved up to now. During final placement's first-fit, we check whether the resulting offset drives the buffer at hand to exceed our record. In that case, we stop, discard the present boxing, and start anew.

We initialize our bootstrapping value with what we consider to be the best heuristic available: sort by size, break ties by lifespan, and do first-fit. A fitting name for it would be ``big-rocks-first''. The bootstrapping value is updated whenever \texttt{idealloc} yields a smaller makespan.

\subsection{Prelude Analysis}
\label{sec:prelude}
Certain tasks need take place only once across \texttt{idealloc}'s flow. Before doing anything else, we bundle the following tasks into a single event traversal: (i) check for elementary cases (Section \ref{sec:elem}), (ii) compute max load, minimum and maximum height, and (iii) construct the interference graph (Section \ref{sec:ig}).

If any of the elementary cases holds, execution proceeds accordingly and an optimal solution is found in minimum time. Else, \texttt{idealloc} must prepare to iterate on its box-and-place core (Sections \ref{sec:coreba}, \ref{sec:unbox}). More specifically:

\begin{itemize}
    \item if the max-to-min height ratio $r$ does not comply with Inequality \ref{eq:rfin}, a ``dummy'' job of height equal to $\lceil 2216.53\cdot h_{min}\rceil$ and lifetime spanning all of the input is added to the buffers to be boxed
    \item bootstrapping takes place as described in Section \ref{sec:bootstr}
    \item the real number $\epsilon$ governing the boxing algorithm is configured as described below
\end{itemize}

Recall that according to Inequality \ref{eq:efin}, it is only within a specific range that $\epsilon$ may move. A simple iterative process is followed to pick the final value: we initialize $\epsilon$ to be equal to the left arm of Ineq. \ref{eq:efin}. We run the boxing algorithm \textit{up to the point where $r^*$ is computed} (see Section \ref{sec:baover}). Next, we increase $\epsilon$ by $1\%$ of the remaining range and repeat. We keep that value which yields the smallest $r^*$.

\subsection{Fast and Correct Final Placement}
\label{sec:ig}
Two extra operations to what was described in Section \ref{sec:unbox} are necessary: if a ``dummy'' job was inserted during prelude analysis, we \textit{ignore} it during unboxing, i.e., we do not assign it any offset and proceed as if it did not exist. Secondly, we ensure that offsets calculated in the final first-fit pass are compliant with each job's potential alignment requirements. Recall that we know both the start address of the range as well as each buffer's alignment (Sections \ref{sec:interface}, \ref{sec:job}).

One further optimization we introduce is an \textit{interference graph}, i.e., a hash map with job IDs as keys, and vectors of concurrently live buffers as values. We use this graph during the first-fit stage, to avoid an otherwise quadratic-complexity overlap check (to be precise, worst-case complexity is still quadratic but in practice rarely does every buffer overlap with everyone else).

\begin{figure}[t!]
    \centering
    \includegraphics[width=\textwidth]{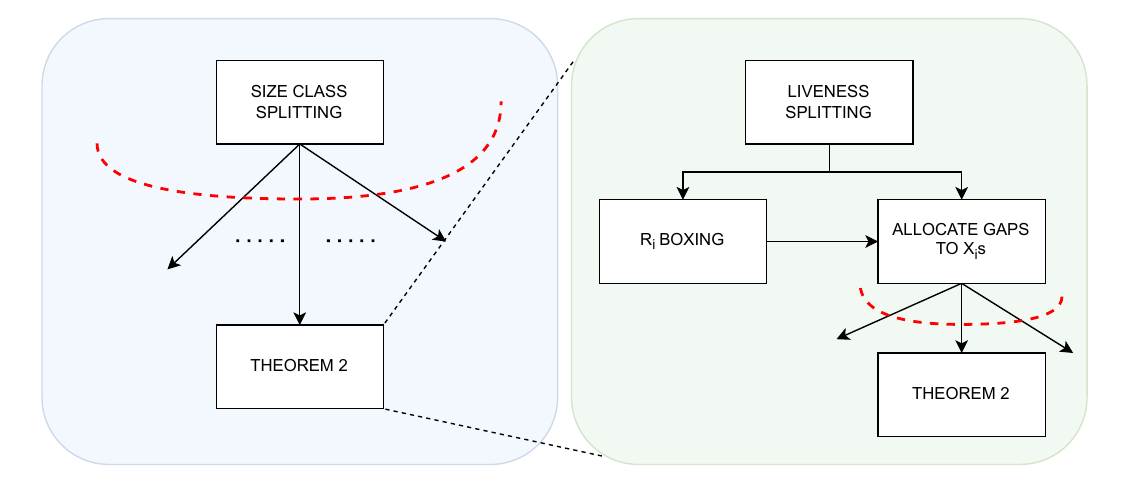}
    \caption{Illustration of parallelism opportunities as thick red dashed curves. The light blue box (left) is Corollary 15. Theorem 2 is invoked on each size class independently. The light green box (right) is a simplified unpacking of Theorem 2. Recursive calls to self are issued for each $X_i$ once all $R_i$s are boxed and gaps shared. Each call is independent from the rest.}
    \label{fig:parbox}
\end{figure}

\subsection{Theorem 2 Simplification}
\label{sec:t2simpl}
The one thing to keep in mind as regards Theorem 2 is that it is expected to box all jobs it is given by Corollary 15 into boxes of size $H$. For reasons tied to their mathematical arguments, Buchsbaum et al. must pretend that first, Corollary 15 scales jobs down to unit height and \textit{then} passes them to Theorem 2 with height parameter $\lfloor H/h\rfloor$, before scaling the returned boxes up back to $H$. \texttt{idealloc} is concrete evidence that the process can both be simplified and remain correct.

The \textit{actual} interface used by Theorem 2 comprises: (i) the set of buffers to be boxed, (ii) the quantity $\lfloor H/h\rfloor$, (iii) box size $H$, (iv) the usual error parameter $\epsilon$, (v) the definition's bounding interval, and (vi) the definition's vector of critical coordinates. There are no ``free spaces'' needed. Boxing happens in two places only: Lemma 1 (see Appendix) and after grouping its unresolved jobs to rows via IGC. As long as \texttt{idealloc} asserts when boxing that the load of the jobs to be boxed does not exceed the expected box height $H$, execution may proceed.

Also note that, when either initializing the critical coordinates vector or injecting points to it as per Section \ref{sec:cpi}, it suffices to consider only those points that appear during event traversal.

\subsection{Parallel Boxing}
There are two opportunities for coarse-grain parallelism in the boxing flow. Both are shown on Figure \ref{fig:parbox}. The first opportunity appears in Corollary 15: the buffers of each size class can be boxed by Theorem 2 independently. The second opportunity appears in Theorem 2, where the recursive calls for each $X_i$ can also be made in parallel. In both cases, no dependencies between parallel tasks exist. We exploit them accordingly to minimize execution time.

\subsection{Doors to Randomness}
\label{sec:random}
Apart from the critical coordinate selection in the context of Theorem 2, there are numerous other spots in our source code that behave non-deterministically \textit{in a baked-in manner}. For instance, there are places where jobs have to be sorted according to some arbitrary criterion, e.g., in reverse de-allocation time. In each such case, again to minimize execution time, we utilize \textit{unstable} sorting, which may re-order equal elements. Another example is the priority queue we are using for event traversal, which does not guarantee that the insertion order of equal elements is preserved.

It is the systemic interaction of all these random effects that gives \texttt{idealloc} its stochasticity.

\begin{table*}
    \caption{Experimental setup used for evaluating \texttt{idealloc}.}
    \label{tab:setup}
    \centering
    \begin{tabular}{||p{0.15\textwidth}|p{0.2\textwidth}|p{0.1\textwidth}|p{0.2\textwidth}|p{0.2\textwidth}||}
        \hline
        \multicolumn{5}{||c||}{\textbf{ALLOCATORS}} \\ \hline
        \textbf{Compiler} & \textbf{Algorithm} & \textbf{Commit} & \multicolumn{2}{|c||}{\textbf{Build Remarks}} \\ \hline
        \multirow{2}{*}{XLA} & Some complex best-fit heuristic. & \multirow{2}{*}{\href{https://github.com/openxla/xla/tree/896c0289645e87e42d2e552c0be2b41d0b886adb}{\texttt{896c02}}} & \multicolumn{2}{|p{0.4\textwidth}||}{\multirow{2}{*}{\texttt{-O3} flag worsened performance.}} \\ \hline
        \multirow{2}{*}{MindSpore} & \multirow{2}{*}{\texttt{SOMAS}~\cite{somas}} & \multirow{2}{*}{\href{https://github.com/mindspore-ai/mindspore/tree/4308a56eab21700459c61db290f47e7e50f4b7f6}{\texttt{4308a56}}} & \multicolumn{2}{|p{0.4\textwidth}||}{CMake build type ``Release'' improved performance, so we kept it.} \\ \hline
        TVM & \texttt{hillclimb} & \href{https://github.com/apache/tvm/tree/cfe1711934f82e56f147f2f5f9f928b5a9b92b3e}{cfe1711} & \multicolumn{2}{|p{0.4\textwidth}||}{Same as \texttt{SOMAS}, Triton.} \\ \hline
        N/A & \texttt{minimalloc}~\cite{minimalloc} & \href{https://github.com/google/minimalloc/tree/987b3c1f9fefe3538ddffa5dc08836831efd3915}{\texttt{987b3c1}} & \multicolumn{2}{|p{0.4\textwidth}||}{None.} \\ \hline
        N/A & \texttt{idealloc} (this paper) & N/A & \multicolumn{2}{|p{0.4\textwidth}||}{Cargo \texttt{--release} flag and LTO enabled.} \\ \hline
        \hline
        \multicolumn{5}{||c||}{\textbf{BENCHMARK SUITES}} \\ \hline
        \multirow{2}{*}{\textbf{Name}} & \multirow{2}{*}{\textbf{Type (Domain)}} & \textbf{\# of Benchmarks} & \textbf{(Smallest, Largest) \# of Buffers} & \multirow{2}{*}{\textbf{Retrieved Via}} \\ \hline
        \multirow{3}{*}{minimalloc} & \multirow{3}{=}{TPU Inference (Unknown)} & \multirow{3}{*}{11} & \multirow{3}{*}{(154, 454)} & minimalloc GitHub repo (``challenging'' suite). \\ \hline
        \multirow{2}{*}{MindSpore} & \multirow{2}{=}{NPU Training (NLP \& Computer Vision)} & \multirow{2}{*}{2} & \multirow{2}{*}{(1042, 18692)} & Emails with the authors of SOMAS~\cite{somas}. \\ \hline
         \multirow{2}{*}{In-house} & \href{https://github.com/google/iopddl/tree/main/benchmarks}{ASPLOS Contest Track}, LevelDB tracing & \multirow{2}{*}{4} & \multirow{2}{*}{(816, 567573)} & \multirow{2}{=}{Custom code.} \\ \hline
    \end{tabular}
\end{table*}

\section{Evaluation}
\label{sec:xps}
We ask the following research questions:

\subsubsection*{\textbf{1. ~Superiority against toy heuristics}}
We have characterized \texttt{idealloc} as a ``stochastic bootstrapped heuristic''. Does it outperform the simplest of heuristics in terms of fragmentation?

\subsubsection*{\textbf{2. ~Degree of randomness}}
Given the high degree of stochasticity elaborated in Section \ref{sec:random}, how probable is that event where applying first-fit to a completely random permutation of the input yields less fragmentation than \texttt{idealloc}?

\subsubsection*{\textbf{3. ~Competence against the SOTA}}
Allocators must (i) produce solutions (ii) of low fragmentation (iii) in reasonable time. We encode this requirement in the following per-benchmark grading scheme: if for \textit{any} reason (e.g., segmentation fault, floating point exception) an allocator crashes, it loses as many points as the allocators that did not. The same if it has not terminated after 15 minutes. In the rest of cases, the allocator earns as many points as the allocators that it outperformed. How many points does \texttt{idealloc} earn under this grading scheme? 

\subsubsection*{\textbf{4. ~Core latency}}
The interface of \texttt{idealloc} exposes the total number of iterations over its box-unbox-place core as a user option (Section \ref{sec:interface}). How cheap is each such iteration?

\subsubsection*{\textbf{5. ~Futureproofness}}
From the outset we have emphasized our interest on DSA instances of arbitrary size and complexity. We want \texttt{idealloc} to fare well against the hardest of possible inputs. If we define hardness as the bootstrap heuristic's fragmentation (we will be calling that heuristic ``SLFF'' from this point onwards), how much better than SLFF is \texttt{idealloc} as hardness grows?\\

The materials used for our experiments are listed in Table \ref{tab:setup}. Note that it was particularly difficult to find non-trivial benchmarks in the sense of SLFF yielding non-zero fragmentation.

\begin{figure}[t!]
    \centering
    \begin{subfigure}[t]{0.33\textwidth}
        \centering
        \includegraphics[width=\textwidth]{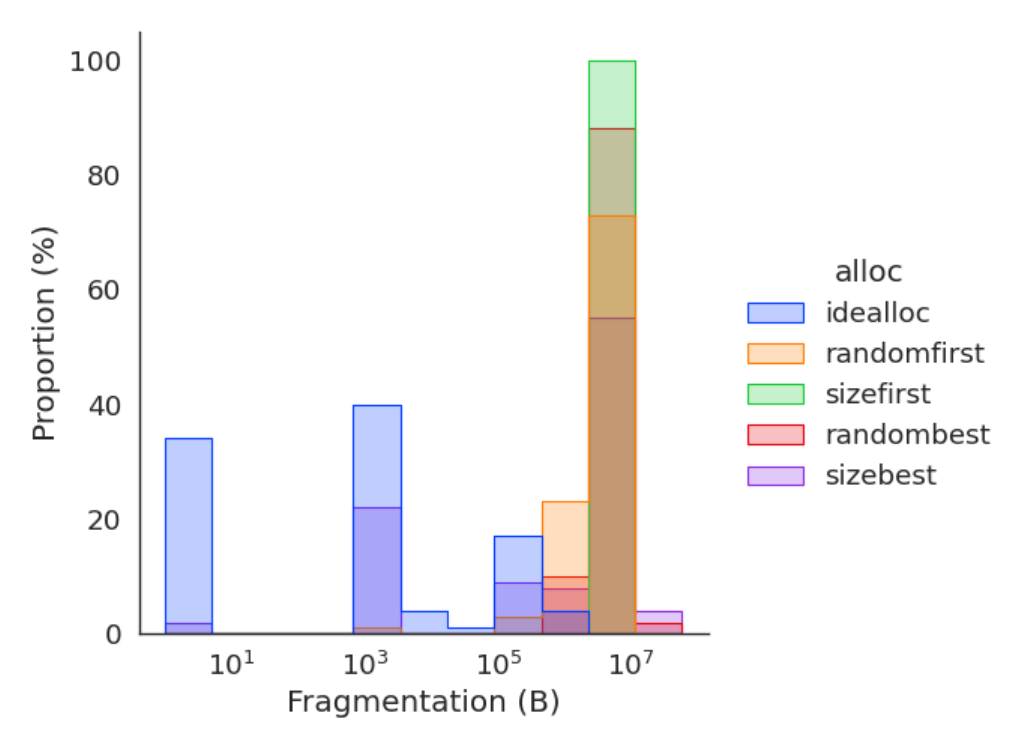}
        \caption{iopddl-G}
    \end{subfigure}
    \hfill
    \begin{subfigure}[t]{0.33\textwidth}
        \centering
        \includegraphics[width=\textwidth]{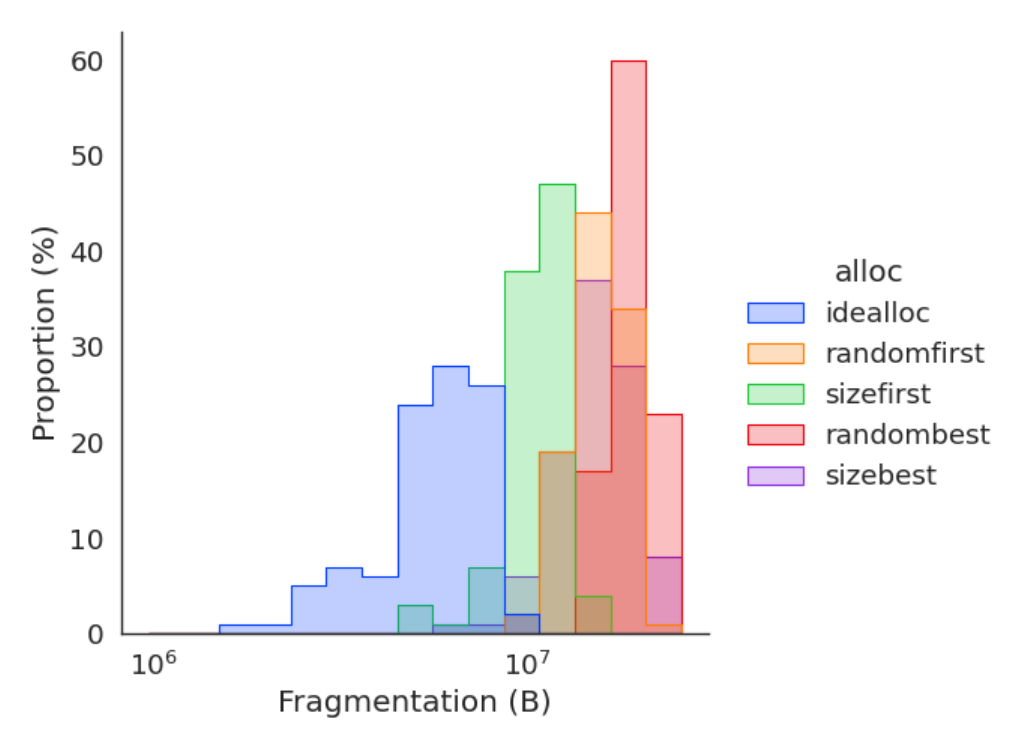}
        \caption{ResNet-50}
    \end{subfigure}
    \hfill
    \begin{subfigure}[t]{0.33\textwidth}
        \centering
        \includegraphics[width=\textwidth]{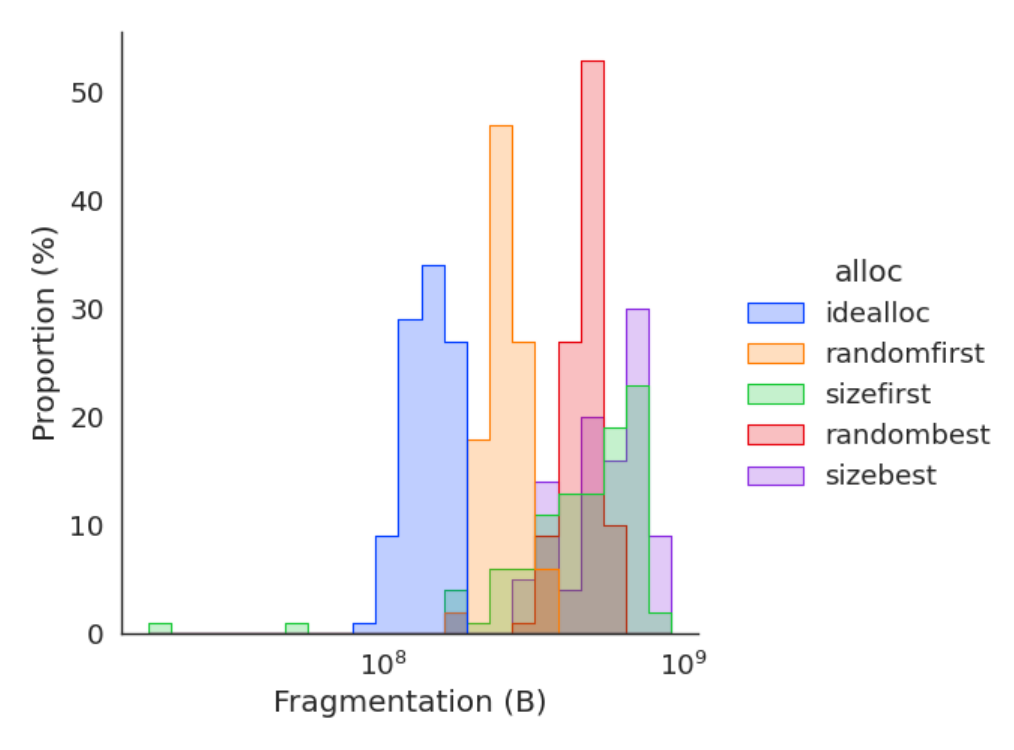}
        \caption{Pangu-2.6B}
    \end{subfigure}
    \begin{subfigure}[t]{0.33\textwidth}
        \centering
        \includegraphics[width=\textwidth]{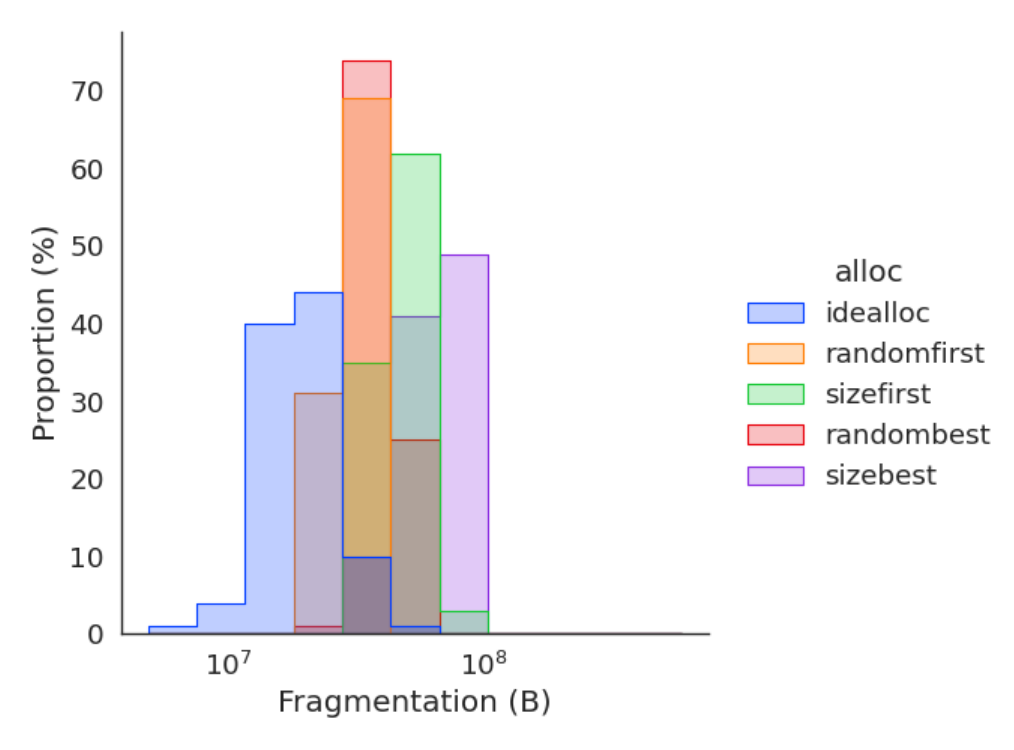}
        \caption{iopddl-S}
    \end{subfigure}
    \hfill
    \begin{subfigure}[t]{0.33\textwidth}
        \centering
        \includegraphics[width=\textwidth]{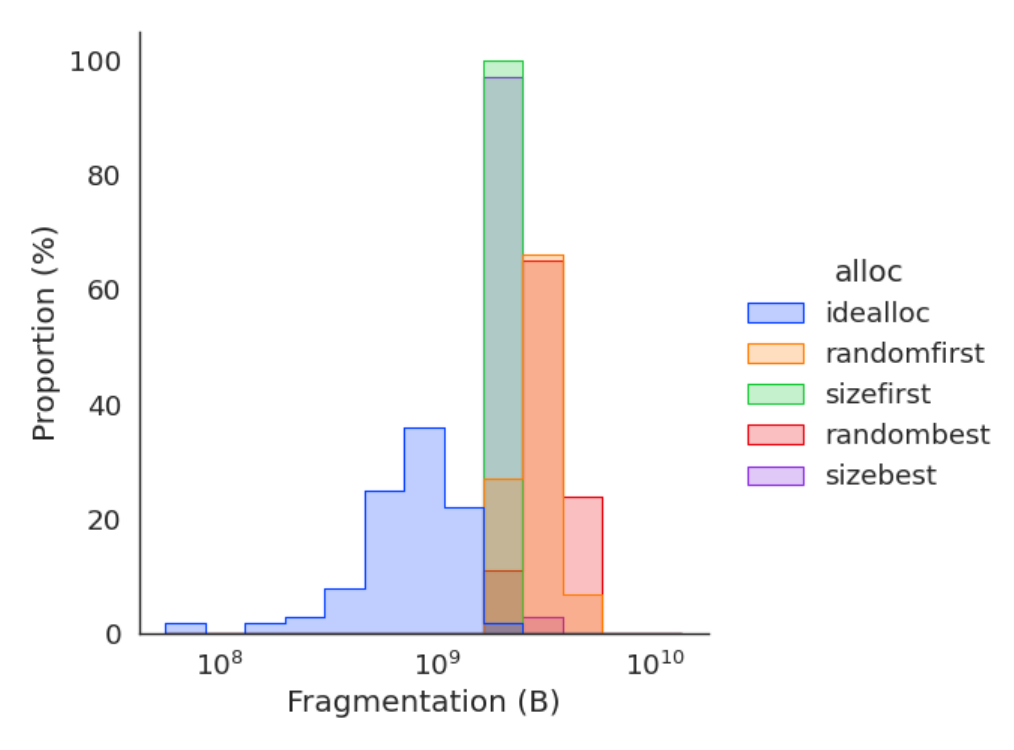}
        \caption{iopddl-Y}
    \end{subfigure}
    \hfill
    \begin{subfigure}[t]{0.33\textwidth}
        \centering
        \includegraphics[width=\textwidth]{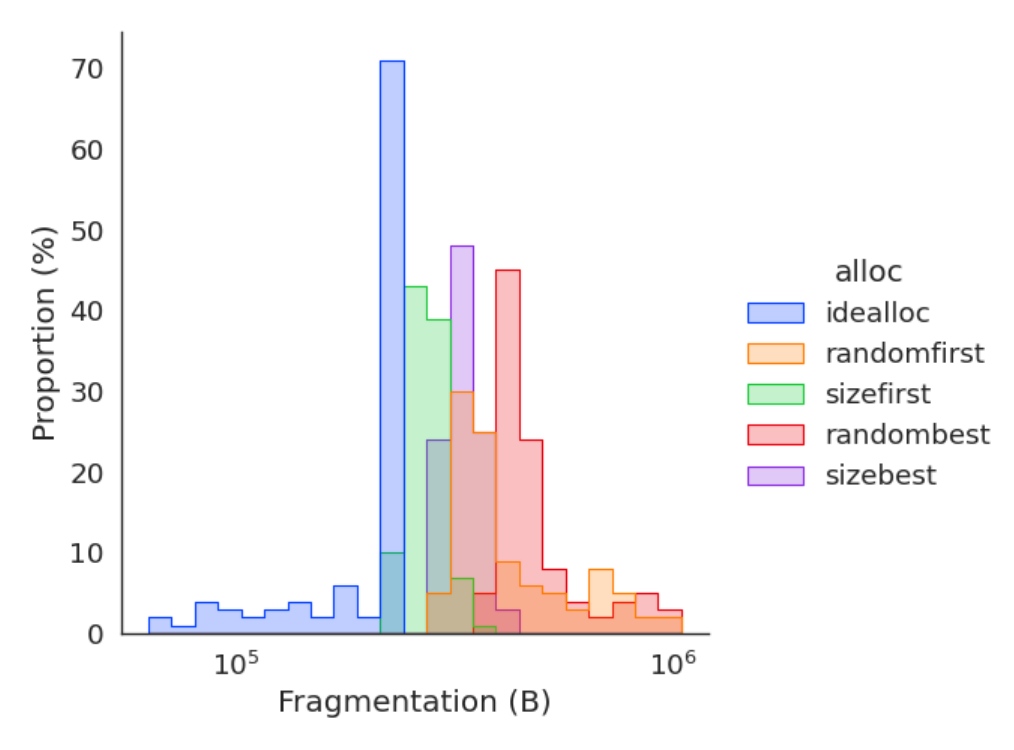}
        \caption{LevelDB}
    \end{subfigure}
    \caption{Fragmentation histograms against heuristics.}
    \label{fig:heurhistos}
\end{figure}

Our measurements took place on a commodity workstation with eight Intel i7-6700 cores clocked at 3.4 GHz, 128 KiB L1 data and instruction caches, 1 MiB L2 and 8 MiB L3. The machine had 32 GiB DRAM and was running Ubuntu 22.04 inside a privileged-mode Docker container. We instrumented all allocators to report allocation time in microseconds excluding I/O. Max memory usage was computed by processing each run's output files and measuring makespan~\footnote{We assume that the target device has no virtual memory and its addresses are physically contiguous. Thus measuring max memory usage offline is accurate. Fellow publications, e.g., minimalloc~\cite{minimalloc}, follow the same practice.}. We executed each benchmark-allocator pair 10 times to ensure statistical integrity. We assigned a maximum allowable time window of 15 minutes per individual run. All measurement scripts were run with a niceness value of $-20$ and minimal background noise.

\begin{figure}[t!]
    \centering
    \begin{subfigure}[t]{0.33\textwidth}
        \centering
        \includegraphics[width=\textwidth]{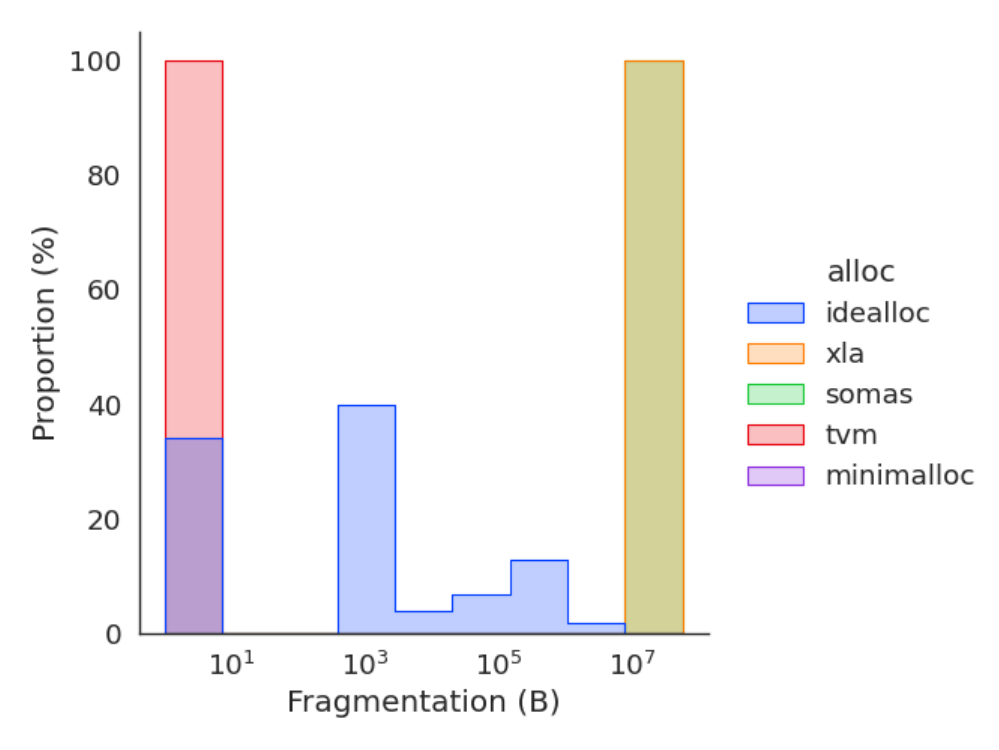}
        \caption{iopddl-G}
    \end{subfigure}
    \hfill
    \begin{subfigure}[t]{0.33\textwidth}
        \centering
        \includegraphics[width=\textwidth]{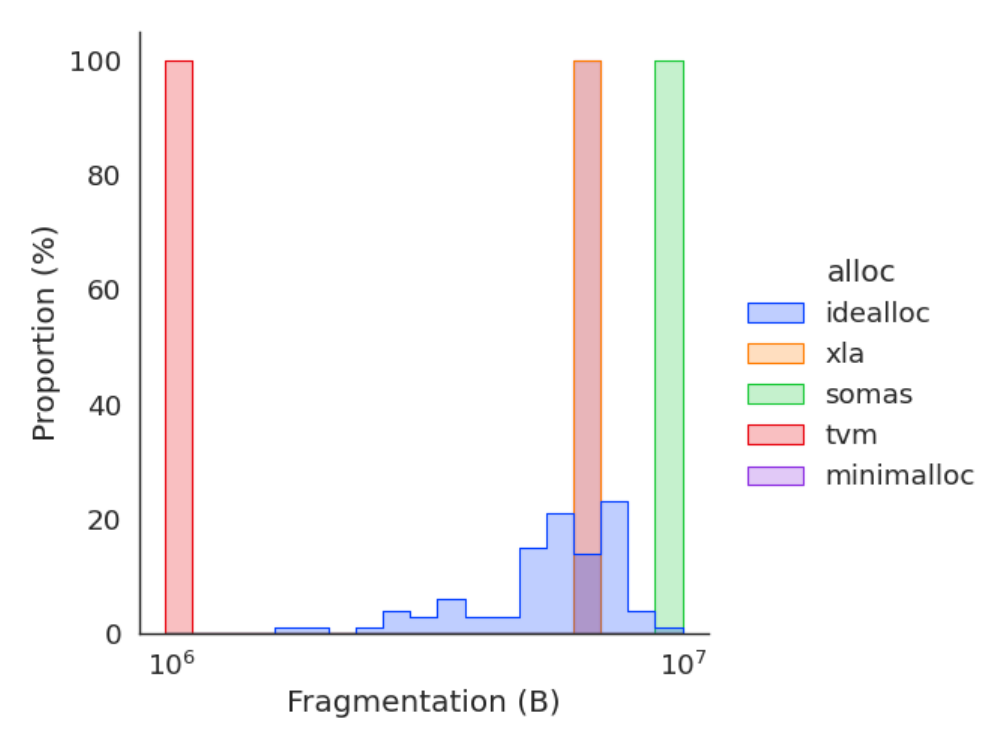}
        \caption{ResNet-50}
    \end{subfigure}
    \hfill
    \begin{subfigure}[t]{0.33\textwidth}
        \centering
        \includegraphics[width=\textwidth]{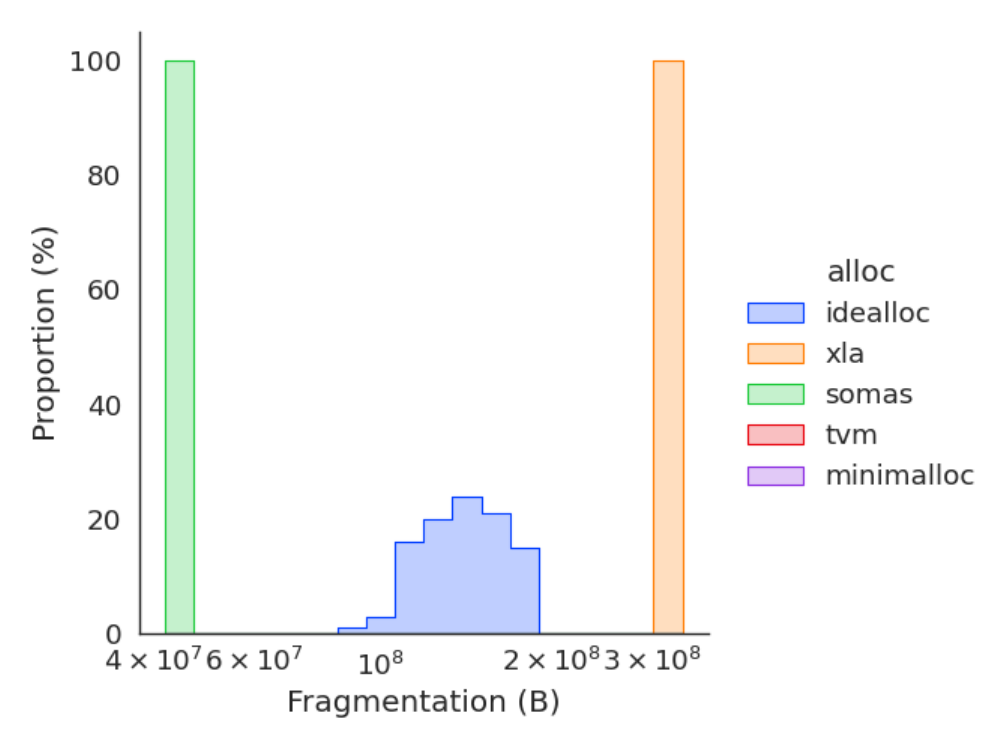}
        \caption{Pangu-2.6B}
    \end{subfigure}
    \begin{subfigure}[t]{0.33\textwidth}
        \centering
        \includegraphics[width=\textwidth]{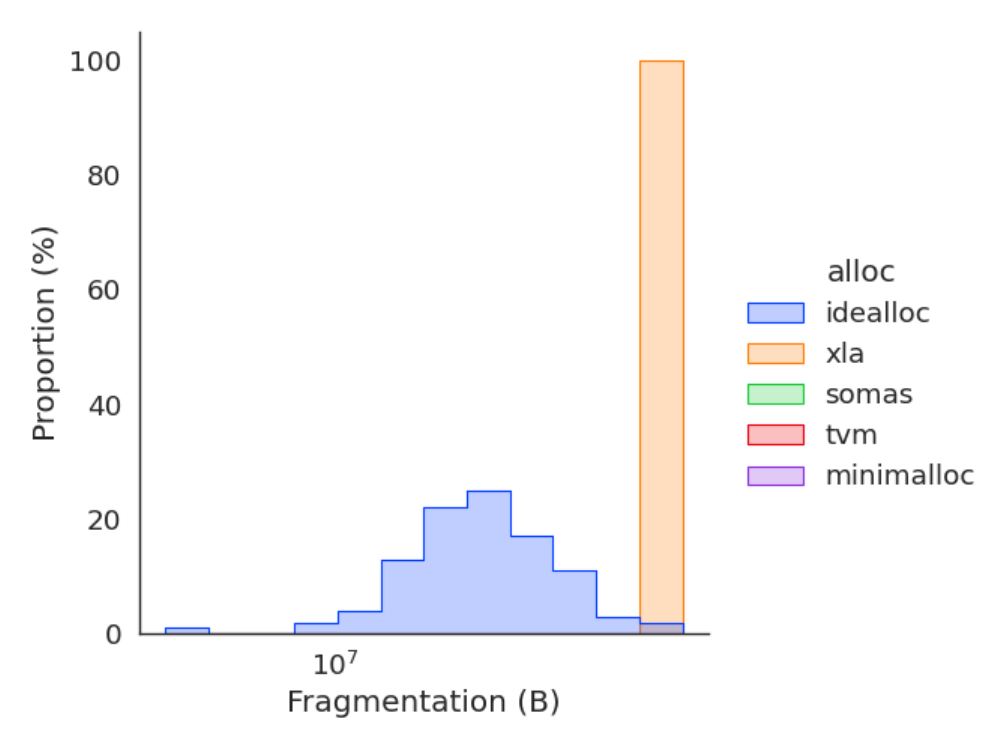}
        \caption{iopddl-S}
    \end{subfigure}
    \hfill
    \begin{subfigure}[t]{0.33\textwidth}
        \centering
        \includegraphics[width=\textwidth]{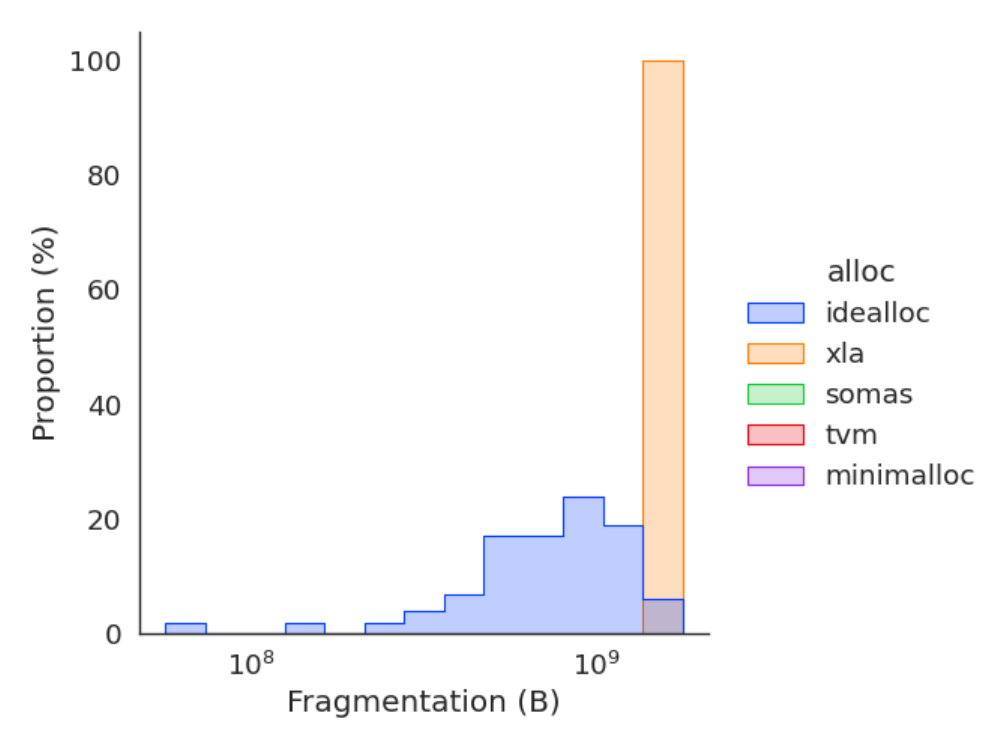}
        \caption{iopddl-Y}
    \end{subfigure}
    \hfill
    \begin{subfigure}[t]{0.33\textwidth}
        \centering
        \includegraphics[width=\textwidth]{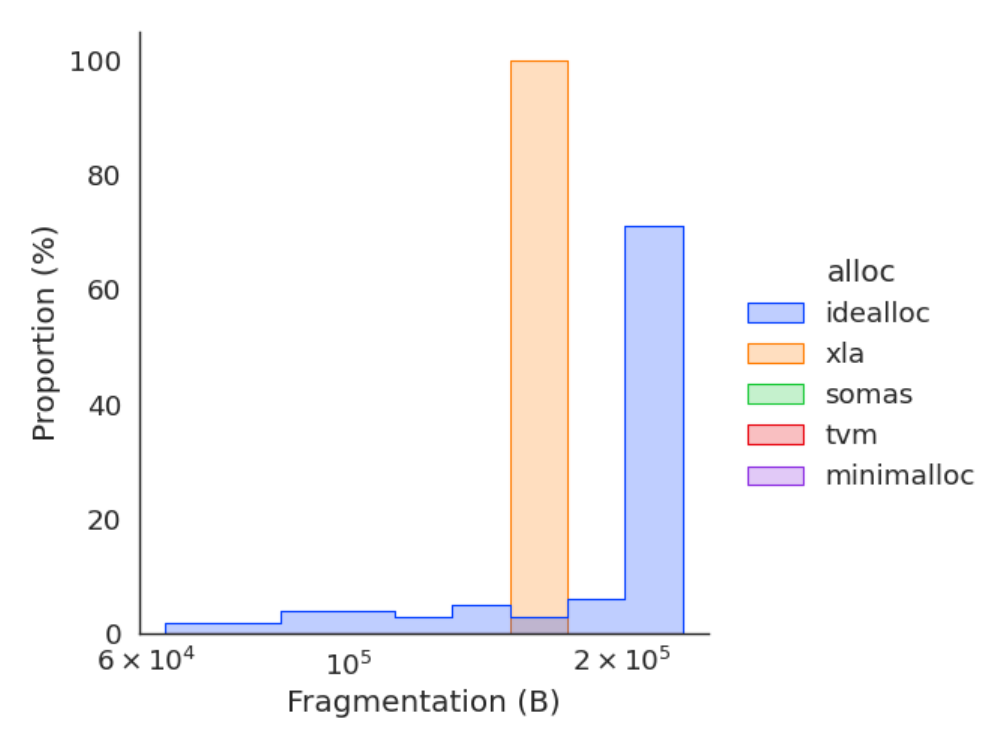}
        \caption{LevelDB}
    \end{subfigure}
    \caption{Fragmentation histograms against the SOTA.}
    \label{fig:histos}
\end{figure}

In addition to the SOTA allocators, we fed each benchmark to \texttt{idealloc} and configured it to run for 100 iterations---except for the LevelDB benchmark, due to whose size we used 10 iterations. In all cases, we repeated our measurements 100 times to let \texttt{idealloc}'s stochasticity express itself as much as possible.

\subsection{Questions 1 and 2}

In Figure \ref{fig:heurhistos} we are comparing \texttt{idealloc}'s fragmentation with four heuristics: the first heuristic (\texttt{sizefirst}) sorts the buffers by decreasing size and then applies first-fit. It is stochastic since, as mentioned, size ties had better be solved at random. The second heuristic (\texttt{randomfirst}) again applies first-fit, but this time on a random permutation of the input buffers. \texttt{sizebest} and \texttt{randombest} are the corresponding best-fit flavors. \texttt{idealloc}'s superiority in all cases is evident.

As a side note, there is no clear indication w.r.t. the superiority of some heuristic over the others. Which one is best, and how they compare to each other varies wildly across benchmarks. Thus using the same heuristic horizontally is guaranteed to waste memory.

\begin{table*}
    \caption{Fragmentation measurements and corresponding points for the \texttt{minimalloc} suite (aggregated) and for the rest of the benchmarks (detailed). As regards marked failures: TVM timed out in all of the benchmarks where it failed. SOMAS timed out in LevelDB and threw a floating point exception in iopddl-S/Y. \texttt{minimalloc} timed out everywhere except iopddl-G, where it segfaulted.}
    \label{tab:inex}
    \centering
    \begin{tabular}{||c|c|c|c|c||}
        \hline
        \textbf{Benchmark~(\#bufs.)} & \textbf{Allocator} & \textbf{Fragmentation} & \textbf{Norm. Frag. (\%)} & \textbf{Points} \\ \hline
        \multirow{5}{*}{\textbf{MINIMALLOC POINTS}} & \texttt{XLA} & \multirow{5}{*}{N/A} & \multirow{5}{*}{N/A} & 1 \\ \cline {2-2} \cline {5-5}
        & \textbf{\texttt{TVM}} & & & \textbf{39} \\ \cline {2-2} \cline {5-5}
        & \texttt{SOMAS} & & & 11 \\ \cline {2-2} \cline {5-5}
        & \texttt{minimalloc} & & & 36 \\ \cline {2-2} \cline {5-5}
        & \texttt{idealloc} & & & 18 \\ \hline
        
        \multirow{5}{*}{iopddl-G~(816)} & \texttt{XLA} & 54.9 MiB & 1.9\% & 1 \\ \cline {2-5}
        & \textbf{\texttt{TVM}} & \textbf{0 MiB} & \textbf{0\%} & \textbf{4} \\ \cline{2-5}
        & \texttt{SOMAS} & 8 MiB & 0.3\% & 2 \\ \cline{2-5}
        & \texttt{minimalloc} & FAILED & FAILED & -4 \\ \cline{2-5}
        & \texttt{idealloc} & 81 KiB & $\sim$0\% & 3 \\ \hline
        
        \multirow{5}{*}{ResNet-50~(1042)} & \texttt{XLA} & 6.4 MiB & 0.45\% & 1 \\ \cline {2-5}
        & \textbf{\texttt{TVM}} & \textbf{946 KiB} & \textbf{0.06\%} & \textbf{4} \\ \cline{2-5}
        & \texttt{SOMAS} & 9.5 MiB & 0.66\% & 0 \\ \cline{2-5}
        & \texttt{minimalloc} & 6.1 MiB & 0.42\% & 2 \\ \cline{2-5}
        & \texttt{idealloc} & 5.5 MiB & 0.38\% & 3 \\ \hline
        
        \multirow{5}{*}{Pangu-2.6B~(18692)} & \texttt{XLA} & 322.8 MiB & 6.1\% & 2 \\ \cline {2-5}
        & \texttt{TVM} & FAILED & FAILED & -3 \\ \cline{2-5}
        & \textbf{\texttt{SOMAS}} & \textbf{40 MiB} & \textbf{0.8\%} & \textbf{4} \\ \cline{2-5}
        & \texttt{minimalloc} & FAILED & FAILED & -3 \\ \cline{2-5}
        & \texttt{idealloc} & 135.2 MiB & 2.6\% & 3 \\ \hline
        
        \multirow{5}{*}{iopddl-S~(28526)} & \texttt{XLA} & 42.5 MiB & 3\% & 3 \\ \cline {2-5}
        & \texttt{TVM} & FAILED & FAILED & -2 \\ \cline{2-5}
        & \texttt{SOMAS} & FAILED & FAILED & -2 \\ \cline{2-5}
        & \texttt{minimalloc} & FAILED & FAILED & -2 \\ \cline{2-5}
        & \textbf{\texttt{idealloc}} & \textbf{18.9 MiB} & \textbf{1.3\%} & \textbf{4} \\ \hline
        
        \multirow{5}{*}{iopddl-Y~(62185)} & \texttt{XLA} & 1.6 GiB & 0.35\% & 3 \\ \cline {2-5}
        & \texttt{TVM} & FAILED & FAILED & -2 \\ \cline{2-5}
        & \texttt{SOMAS} & FAILED & FAILED & -2 \\ \cline{2-5}
        & \texttt{minimalloc} & FAILED & FAILED & -2 \\ \cline{2-5}
        & \textbf{\texttt{idealloc}} & \textbf{771.7 MiB} & \textbf{0.16\%} & \textbf{4} \\ \hline
        
        \multirow{5}{*}{LevelDB~(567573)} & \textbf{\texttt{XLA}} & \textbf{160 KiB} & \textbf{0.6\%} & \textbf{4} \\ \cline {2-5}
        & \texttt{TVM} & FAILED & FAILED & -2 \\ \cline{2-5}
        & \texttt{SOMAS} & FAILED & FAILED & -2 \\ \cline{2-5}
        & \texttt{minimalloc} & FAILED & FAILED & -2 \\ \cline{2-5}
        & \texttt{idealloc} & 198 KiB & 0.8\% & 3 \\ \hline

        \multirow{5}{*}{\textbf{REST POINTS}} & \texttt{XLA} & \multirow{5}{*}{N/A} & \multirow{5}{*}{N/A} & 14 \\ \cline {2-2} \cline {5-5}
        & \texttt{TVM} & & & -1 \\ \cline {2-2} \cline {5-5}
        & \texttt{SOMAS} & & & 0 \\ \cline {2-2} \cline {5-5}
        & \texttt{minimalloc} & & & -11 \\ \cline {2-2} \cline {5-5}
        & \textbf{\texttt{idealloc}} & & & \textbf{20} \\ \hline
        
        \multirow{5}{*}{\textbf{TOTAL POINTS}} & \texttt{XLA} & \multirow{5}{*}{N/A} & \multirow{5}{*}{N/A} & 15 \\ \cline {2-2} \cline {5-5}
        & \textbf{\texttt{TVM}} & & & \textbf{38} \\ \cline {2-2} \cline {5-5}
        & \texttt{SOMAS} & & & 11 \\ \cline {2-2} \cline {5-5}
        & \texttt{minimalloc} & & & 25 \\ \cline {2-2} \cline {5-5}
        & \textbf{\texttt{idealloc}} & & & \textbf{38} \\ \hline
    \end{tabular}
\end{table*}

\subsection{Question 3}

From the opponent allocators, XLA uses \texttt{In} semantics, and \texttt{minimalloc}, SOMAS and TVM use \texttt{InEx}. \texttt{idealloc}, on the other hand, uses \texttt{Ex}. To ensure fairness we conducted the analysis described in Section \ref{sec:sem} and decided to assume that all of our benchmarks use \texttt{InEx} semantics. We then took our measurements and plotted fragmentation histograms like the ones shown in Figure \ref{fig:histos}. The respective rankings are listed in Table \ref{tab:inex}. The same table includes a summary of the rankings formed for the \texttt{minimalloc} micro-benchmarks.

\subsection{Question 4}

\begin{wrapfigure}[17]{r}{0.55\textwidth}
    \vspace{-0.5cm}
    \begin{minipage}{0.55\textwidth}
        \begin{figure}[H]
            \centering
            \includegraphics[width=\textwidth]{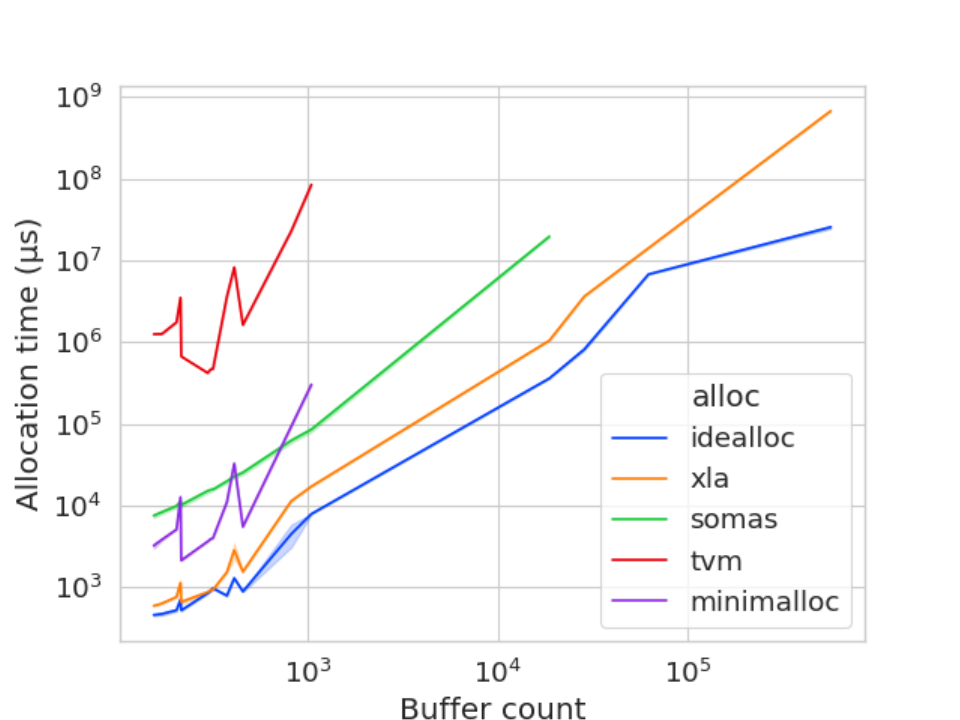}
        \end{figure}       
    \end{minipage}
    \caption{\texttt{idealloc}'s single-iteration latency versus its competition, as a function of total buffer count. Note the interference graph's impact at the far end of the curve.}
    \label{fig:inexlat}
\end{wrapfigure}

We plot allocation time as a function of the buffer count in Figure \ref{fig:inexlat}. Particularly w.r.t. \texttt{idealloc} we have plotted \textit{single-iteration} latency, which includes one prelude analysis (Section \ref{sec:prelude}) and a single box-unbox-place pass (Sections \ref{sec:coreba}, \ref{sec:unbox}). Regardless from the size of the input, \texttt{idealloc}'s core latency is faster than any alternative.

\subsection{Question 5}

\begin{wrapfigure}[15]{r}{0.55\textwidth}
    \vspace{-1.1cm}
    \begin{minipage}{0.55\textwidth}
        \begin{figure}[H]
            \centering
            \includegraphics[width=\textwidth]{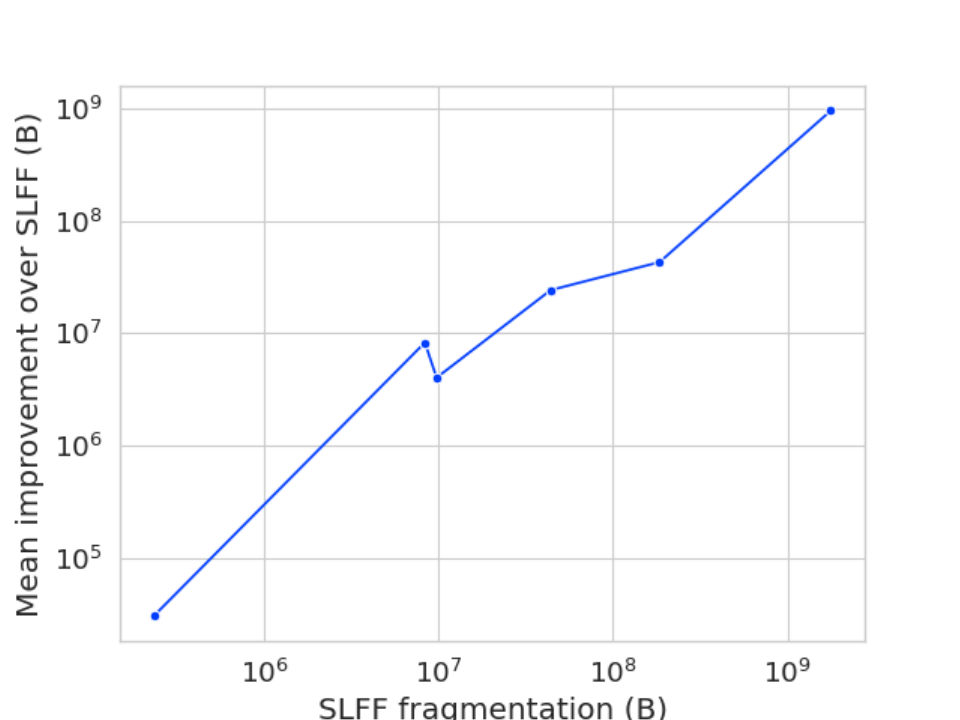}
        \end{figure}       
    \end{minipage}
    \caption{\texttt{idealloc}'s mean improvement over its bootstrap heuristic as a function of the bootstrap heuristic's own fragmentation.}
    \label{fig:antifragile}
\end{wrapfigure}

We see in Figure \ref{fig:antifragile} that \texttt{idealloc} outperforms SLFF in a steady fashion as the input hardness increases. The ideal but impossible scenario would be for the drawn line to coincide with $y=x$, i.e., for boxing to always eliminate fragmentation completely. It nevertheless stays close enough.

\section{Discussion}
\label{sec:disc}
We began our exposition by declaring our interest in real allocators and how they behave under pressure. We have now presented evidence that (i) there is a gap in the SOTA as regards effective \textit{and} scalable solutions, and (ii) \texttt{idealloc} fills that gap. That said, we are aware of the subtleties involved in the process toward making such strong statements. The first half of this Section examines said subtleties from close distance. We then discuss meaningful future activities to either improve or utilize our allocator.

\subsection{Results and Their Interpretation}
An important point to agree on is whether the selected allocators listed in Table \ref{tab:setup} reflect what we mean by ``DSA SOTA''. Our initial measurements also included three greedy algorithms from LiteRT (formerly TensorFlow Lite)~\cite{pisarchyk2020efficient} and one from OpenAI's Triton~\cite{triton}. Furthermore, XLA features a second allocator based on heap simulation\footnote{\href{https://github.com/openxla/xla/blob/main/xla/service/heap_simulator/heap_simulator.h}{https://github.com/openxla/xla/blob/main/xla/service/heap\_simulator/heap\_simulator.h}}, mimicking an on-line OS allocator. TVM has heuristics similar to \texttt{sizefirst} besides the hillclimb algorithm\footnote{\href{https://github.com/apache/tvm/blob/cfe1711934f82e56f147f2f5f9f928b5a9b92b3e/src/tir/usmp/algo/greedy.cc}{https://github.com/apache/tvm/blob/cfe1711934f82e56f147f2f5f9f928b5a9b92b3e/src/tir/usmp/algo/greedy.cc}}. IREE's one and only algorithm is a sort-by-allocation-time best-fit heuristic\footnote{\href{https://github.com/iree-org/iree/blob/15ca58e19ec76fab94c4aba8f75091c532282d51/compiler/src/iree/compiler/Dialect/Stream/Transforms/LayoutSlices.cpp}{https://github.com/iree-org/iree/blob/15ca58e19ec76fab94c4aba8f75091c532282d51/compiler/src/iree/compiler/Dialect/Stream/Transforms/LayoutSlices.cpp}}. We included all these as well, but their performance was poor and we decided to keep our tables and figures from getting too crowded. The only ``popular'' deep learning compiler we did not inspect was Meta's Glow, an omission owed to lack of time, not of meticulousness. ILP formulations of DSA are known to be inferior due to poor scalability~\cite{telamalloc,minimalloc}. We thus believe to have cast a wide and informed enough gaze.

Let us now visit some more specific issues:

\subsubsection{Grading System Fairness}
Since the crux of our argument is the rankings of Table \ref{tab:inex}, asking if our grading system is fair is a fair question. We used a \textit{tournament} comprising many \textit{races} as a model. The results of each race, i.e., benchmark, are translated to points for each allocator. Whoever has collected the most points after the last race is the tournament's winner. This model is fair to the extent that the points translation scheme is.

Our scheme \textit{rewards} allocators with as many points as the allocators they beat. The number includes both those that yielded worse fragmentation and those that failed. The only objection we can think of is that \textit{differences} in fragmentation are not accounted for. However, the same holds in an actual racing tournament: individual times don't matter.

Moreover, our scheme \textit{punishes} failing allocators with as many points as the allocators that did not fail. Why did we not use a fixed punishment, i.e., losing one point at each failure? Imagine a tournament of $N$ contestants. Focus on contestants $A$ and $B$. In the first race of the tournament, $A$ finishes first and $B$ is the only contestant that did not finish at all. In the second race, $B$ is the only finisher. Under a fixed-punishment scheme, $A$ and $B$ would end up with $N-2$ points. Under our scheme, the respective points would be $N-2$ and \textit{zero}. Which one is fairest?

It depends on the type of tournament winner we are searching for. Since almost everyone finished it, the first race of our example was rather easy (think about iopddl-G in Table \ref{tab:inex}). The converse holds for the second race (think LevelDB). So do we want to incentivize ``laziness'' in easy races for the sake of potential triumph in hard ones? To the authors of this paper, a positive answer sounds like gambling.

\subsubsection{On Normalized Fragmentation}
Despite including normalized values for fragmentation in Table \ref{tab:inex}, i.e., absolute fragmentation divided by the benchmark's max load, we do not encourage their use. In the age of ``memory walls''~\cite{ai_memory_wall} memory savings are valuable regardless from necessary memory investment. The reason is simple: most of the time, memory is shared. Savings that look insignificant in proportion to max load can still be used to host data that is foreign to the problem at hand. Only when considering things \textit{in isolation} do absolute quantities lose their weight.

If the above was not convincing enough, consider that by relying on normalized fragmentation, wasting 1 KiB under a max load of 10 KiB looks identical to wasting 1 GiB under a max load of 10 GiB. Both cases have 10\% normalized fragmentation, but the second case is clearly more damaging.

\subsubsection{Core vs. Total Latency}
\begin{wrapfigure}[15]{r}{0.55\textwidth}
    \vspace{-0.5cm}
    \begin{minipage}{0.55\textwidth}
        \begin{figure}[H]
            \centering
            \includegraphics[width=\textwidth]{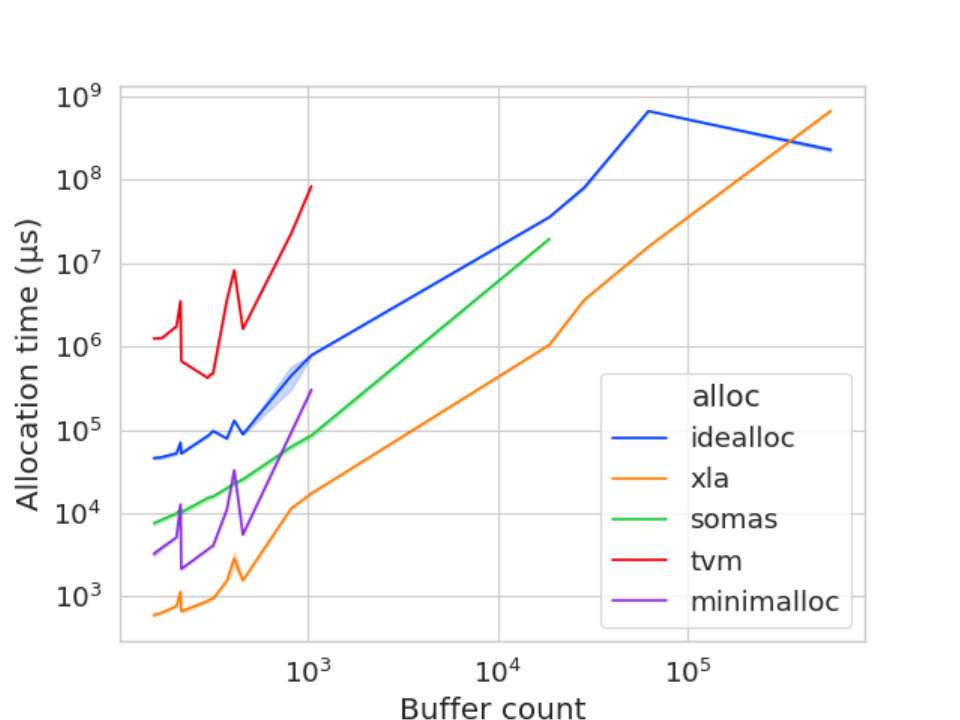}
        \end{figure}       
    \end{minipage}
    \caption{\texttt{idealloc}'s total latency versus its competition.}
    \label{fig:totlat}
\end{wrapfigure}

As noted by Figure \ref{fig:inexlat}'s caption, the plotted blue line stands for \texttt{idealloc}'s \textit{single-iteration} latency. For LevelDB, however, we configured \texttt{idealloc} to repeat 10 iterations, and for the rest of the benchmarks 100. The following remarks apply:

\begin{itemize}
    \item our intention was to highlight the fact that each \texttt{idealloc} iteration takes minimum time compared to the SOTA
    
    \item even when scaled to its real latency (Figure \ref{fig:totlat}), \texttt{idealloc} (i) is up to two orders of magnitude faster than TVM, and (ii) ends up faster than XLA in LevelDB's context

    \item if total allocation time is the user's main concern, off-the-shelf heuristics are the way to go. Otherwise trading off latency for lowering fragmentation stands to reason
\end{itemize}

\subsubsection{Hardness Definition}
While forming our research questions for Section \ref{sec:xps}, we did not explain our decision to define hardness as SLFF's fragmentation. We hope to give a convincing answer here.

The context was that of ``futureproofness'': that arbitrarily hard DSA instances will emerge was, as stated in Section \ref{sec:intro}, our founding assumption. Our prime interest is to ensure that \texttt{idealloc} will be able to deal with them. The hardness we have in mind concerns the \textit{topology} of an instance, that is, the complexity of the landscape formed by the co-existence of a given set of buffer conflicts and the corresponding buffer sizes. For example, an instance where all buffers overlap is not at all hard/complex/non-trivial: even bump allocation would yield zero fragmentation!

We posit that \textit{a reasonable way to gauge the hardness of an instance is to measure the fragmentation incurred by a simple yet decent heuristic}. Consider the problem of packing one's suitcase before a long trip: does it not make sense to start with the biggest of items, and work our way down? If we place all of our items in this fashion, our baggage was not hard to treat. If on the other hand the ``big-rocks-first'' strategy fails, our baggage is as hard as the total size of items that we were forced to leave out. Choosing SLFF as our hardness measure is the DSA equivalent of what we described.

\subsection{Proposed Future Work}

\subsubsection{Sampling Many $\epsilon$-values}
\texttt{idealloc}'s boxing core is governed by the error parameter $\epsilon$ (Section \ref{sec:baover}), which must be confined into an input-specific range of values (Section \ref{sec:latent}). Our current approach is to conduct a preprocessing step where we iterate on the aforementioned range and finally set $\epsilon$ to that value which minimizes the resulting boxing's max-to-min height ratio (Section \ref{sec:prelude}). There is no concrete reason behind this strategy, only the intuition that deeper boxing recursions lead to lower fragmentation. A promising alternative would be to sample $\epsilon$ at random, thus eliminating significant overhead from our prelude analysis and adding more variety to the placements explored.

\subsubsection{Statistical Inference}
Given \texttt{idealloc}'s stochastic nature (Section \ref{sec:random}), it would be good to have an estimate of how many iterations are needed to become certain that most of the solution space has been explored. This implies a statistical inference component observing each iteration's makespan and using it to refine an on-line distribution. However, such observability would incur performance overhead: to monitor all makespans we would have to remove early stopping (Section \ref{sec:bootstr}). Moreover, extra time would be needed for ste statistical inference core itself.

\subsubsection{Randomness Taming}
It is tempting to think of some meta-optimization over (i) the selection of $\epsilon$ (Sections \ref{sec:baover}, \ref{sec:bootstr}) and (ii) Theorem 2 critical points (Section \ref{sec:cpi}). This would help us avoid ``useless'' iterations. The main problem with setting up such a mechanism is that our current implementation has non-deterministic elements that are outside our control (Section \ref{sec:random}). On top of that, meta-optimization would need to keep and act on some global state, which would need to be synchronized between threads. In turn this would make everything slower.

\subsection{A Note on Time and Space}
For the entirety of this text we have been interpreting the horizontal dimension as ``time'' and the vertical one as ``space''. We owe this to the fact that our research has its roots in computer systems' memory allocation. Nevertheless, other interpretations could enable using \texttt{idealloc} (or any similar piece of related work) in completely different contexts. For instance, one could view the horizontal axis as a spectrum of frequencies, and the vertical one as time. Each ``buffer'' could thus encode a radio host's request to broadcast over a specific band of frequencies for a specific amount of time. Solving DSA in that context would ensure that (i) all hosts receive a slot for their show and (ii) the overall spectrum is ``reserved'' for as little time as possible.

Room for nuance exists even within the standard time/space interpretation. Whether the vertical axis stands for physical or virtual addresses is left to the hands of the end user. Whether time is wall clock time or, e.g., the total number of bytes allocated by a program, or the indices of a topologically sorted computation graph's nodes, again this decision belongs to the user. DSA itself is indifferent to these decisions. In order for its output to be useful, however, the following invariants must hold: (i) both dimensions must be \textit{contiguous}, i.e., elements that overlap in one dimension cannot do so in the other and (ii) elements are \textit{fixed} in one dimension, and are allowed to ``slide'' only along the other.

\section{Conclusion}
\label{sec:end}
Static memory planning is an NP-complete problem with applications of great potential. Existing solutions are either scalable or memory-efficient. We have presented \texttt{idealloc}, an implementation designed with low fragmentation, high performance and scalability in mind. Along the way we have reported numerous insights that may prove useful to practitioners and theorists in the future.

We have open-source \texttt{idealloc} and the benchmarks used~\footnote{\href{https://github.com/cappadokes/idealloc}{https://github.com/cappadokes/idealloc}}.

\begin{acks}
This work has been supported with funding from the European Union's Horizon research and innovation programme under grant agreement No. 101070374.
\end{acks}

\bibliographystyle{ACM-Reference-Format}
\bibliography{sample-base}


\begin{thebibliography}{29}


\ifx \showCODEN    \undefined \def \showCODEN     #1{\unskip}     \fi
\ifx \showDOI      \undefined \def \showDOI       #1{#1}\fi
\ifx \showISBNx    \undefined \def \showISBNx     #1{\unskip}     \fi
\ifx \showISBNxiii \undefined \def \showISBNxiii  #1{\unskip}     \fi
\ifx \showISSN     \undefined \def \showISSN      #1{\unskip}     \fi
\ifx \showLCCN     \undefined \def \showLCCN      #1{\unskip}     \fi
\ifx \shownote     \undefined \def \shownote      #1{#1}          \fi
\ifx \showarticletitle \undefined \def \showarticletitle #1{#1}   \fi
\ifx \showURL      \undefined \def \showURL       {\relax}        \fi
\providecommand\bibfield[2]{#2}
\providecommand\bibinfo[2]{#2}
\providecommand\natexlab[1]{#1}
\providecommand\showeprint[2][]{arXiv:#2}

\bibitem[Agrawal et~al\mbox{.}(2023)]%
        {footprint}
\bibfield{author}{\bibinfo{person}{Sakshi Agrawal}, \bibinfo{person}{Priyankar Ghosh}, \bibinfo{person}{Gaurav Kumar}, {and} \bibinfo{person}{Tripuraneni Radhika}.} \bibinfo{year}{2023}\natexlab{}.
\newblock \showarticletitle{Memory Footprint Optimization for Neural Network Inference in Mobile SoCs}. In \bibinfo{booktitle}{\emph{2023 IEEE Women in Technology Conference (WINTECHCON)}}. \bibinfo{pages}{1--6}.
\newblock
\urldef\tempurl%
\url{https://doi.org/10.1109/WINTECHCON58518.2023.10277304}
\showDOI{\tempurl}


\bibitem[Buchsbaum et~al\mbox{.}(2003)]%
        {buchsbaum}
\bibfield{author}{\bibinfo{person}{Adam~L. Buchsbaum}, \bibinfo{person}{Howard Karloff}, \bibinfo{person}{Claire Kenyon}, \bibinfo{person}{Nick Reingold}, {and} \bibinfo{person}{Mikkel Thorup}.} \bibinfo{year}{2003}\natexlab{}.
\newblock \showarticletitle{OPT versus LOAD in dynamic storage allocation}. In \bibinfo{booktitle}{\emph{Proceedings of the Thirty-Fifth Annual ACM Symposium on Theory of Computing}} (San Diego, CA, USA) \emph{(\bibinfo{series}{STOC '03})}. \bibinfo{publisher}{Association for Computing Machinery}, \bibinfo{address}{New York, NY, USA}, \bibinfo{pages}{556–564}.
\newblock
\showISBNx{1581136749}
\urldef\tempurl%
\url{https://doi.org/10.1145/780542.780624}
\showDOI{\tempurl}


\bibitem[Garey and Johnson(1979)]%
        {garey1979computers}
\bibfield{author}{\bibinfo{person}{Michael~R Garey} {and} \bibinfo{person}{David~S Johnson}.} \bibinfo{year}{1979}\natexlab{}.
\newblock \bibinfo{booktitle}{\emph{Computers and intractability}}. Vol.~\bibinfo{volume}{174}.
\newblock \bibinfo{publisher}{freeman San Francisco}.
\newblock


\bibitem[Gergov(1996)]%
        {gergov_1}
\bibfield{author}{\bibinfo{person}{Jordan Gergov}.} \bibinfo{year}{1996}\natexlab{}.
\newblock \showarticletitle{Approximation algorithms for dynamic storage allocation}. In \bibinfo{booktitle}{\emph{Algorithms --- ESA '96}}, \bibfield{editor}{\bibinfo{person}{Josep Diaz} {and} \bibinfo{person}{Maria Serna}} (Eds.). \bibinfo{publisher}{Springer Berlin Heidelberg}, \bibinfo{address}{Berlin, Heidelberg}, \bibinfo{pages}{52--61}.
\newblock
\showISBNx{978-3-540-70667-0}


\bibitem[Gergov(1999)]%
        {gergov_2}
\bibfield{author}{\bibinfo{person}{Jordan Gergov}.} \bibinfo{year}{1999}\natexlab{}.
\newblock \showarticletitle{Algorithms for compile-time memory optimization}. In \bibinfo{booktitle}{\emph{Proceedings of the Tenth Annual ACM-SIAM Symposium on Discrete Algorithms}} (Baltimore, Maryland, USA) \emph{(\bibinfo{series}{SODA '99})}. \bibinfo{publisher}{Society for Industrial and Applied Mathematics}, \bibinfo{address}{USA}, \bibinfo{pages}{907–908}.
\newblock
\showISBNx{0898714346}


\bibitem[Gholami et~al\mbox{.}(2024)]%
        {ai_memory_wall}
\bibfield{author}{\bibinfo{person}{Amir Gholami}, \bibinfo{person}{Zhewei Yao}, \bibinfo{person}{Sehoon Kim}, \bibinfo{person}{Coleman Hooper}, \bibinfo{person}{Michael~W. Mahoney}, {and} \bibinfo{person}{Kurt Keutzer}.} \bibinfo{year}{2024}\natexlab{}.
\newblock \showarticletitle{AI and Memory Wall}.
\newblock \bibinfo{journal}{\emph{IEEE Micro}} \bibinfo{volume}{44}, \bibinfo{number}{3} (\bibinfo{year}{2024}), \bibinfo{pages}{33--39}.
\newblock
\urldef\tempurl%
\url{https://doi.org/10.1109/MM.2024.3373763}
\showDOI{\tempurl}


\bibitem[Guo et~al\mbox{.}(2024)]%
        {gmlake}
\bibfield{author}{\bibinfo{person}{Cong Guo}, \bibinfo{person}{Rui Zhang}, \bibinfo{person}{Jiale Xu}, \bibinfo{person}{Jingwen Leng}, \bibinfo{person}{Zihan Liu}, \bibinfo{person}{Ziyu Huang}, \bibinfo{person}{Minyi Guo}, \bibinfo{person}{Hao Wu}, \bibinfo{person}{Shouren Zhao}, \bibinfo{person}{Junping Zhao}, {and} \bibinfo{person}{Ke Zhang}.} \bibinfo{year}{2024}\natexlab{}.
\newblock \showarticletitle{GMLake: Efficient and Transparent GPU Memory Defragmentation for Large-scale DNN Training with Virtual Memory Stitching}. In \bibinfo{booktitle}{\emph{Proceedings of the 29th ACM International Conference on Architectural Support for Programming Languages and Operating Systems, Volume 2}} (La Jolla, CA, USA) \emph{(\bibinfo{series}{ASPLOS '24})}. \bibinfo{publisher}{Association for Computing Machinery}, \bibinfo{address}{New York, NY, USA}, \bibinfo{pages}{450–466}.
\newblock
\showISBNx{9798400703850}
\urldef\tempurl%
\url{https://doi.org/10.1145/3620665.3640423}
\showDOI{\tempurl}


\bibitem[Imanishi and Xu(2024)]%
        {ismm2}
\bibfield{author}{\bibinfo{person}{Akifumi Imanishi} {and} \bibinfo{person}{Zijian Xu}.} \bibinfo{year}{2024}\natexlab{}.
\newblock \showarticletitle{A Heuristic for Periodic Memory Allocation with Little Fragmentation to Train Neural Networks}. In \bibinfo{booktitle}{\emph{Proceedings of the 2024 ACM SIGPLAN International Symposium on Memory Management}} (Copenhagen, Denmark) \emph{(\bibinfo{series}{ISMM 2024})}. \bibinfo{publisher}{Association for Computing Machinery}, \bibinfo{address}{New York, NY, USA}, \bibinfo{pages}{82–94}.
\newblock
\showISBNx{9798400706158}
\urldef\tempurl%
\url{https://doi.org/10.1145/3652024.3665508}
\showDOI{\tempurl}


\bibitem[Kierstead(1991)]%
        {KIERSTEAD1991231}
\bibfield{author}{\bibinfo{person}{H.A. Kierstead}.} \bibinfo{year}{1991}\natexlab{}.
\newblock \showarticletitle{A polynomial time approximation algorithm for dynamic storage allocation}.
\newblock \bibinfo{journal}{\emph{Discrete Mathematics}} \bibinfo{volume}{88}, \bibinfo{number}{2} (\bibinfo{year}{1991}), \bibinfo{pages}{231--237}.
\newblock
\showISSN{0012-365X}
\urldef\tempurl%
\url{https://doi.org/10.1016/0012-365X(91)90011-P}
\showDOI{\tempurl}


\bibitem[Kierstead(1988)]%
        {kierstead_1}
\bibfield{author}{\bibinfo{person}{H.~A. Kierstead}.} \bibinfo{year}{1988}\natexlab{}.
\newblock \showarticletitle{The Linearity of First-Fit Coloring of Interval Graphs}.
\newblock \bibinfo{journal}{\emph{SIAM Journal on Discrete Mathematics}} \bibinfo{volume}{1}, \bibinfo{number}{4} (\bibinfo{year}{1988}), \bibinfo{pages}{526--530}.
\newblock
\urldef\tempurl%
\url{https://doi.org/10.1137/0401048}
\showDOI{\tempurl}
\showeprint{https://doi.org/10.1137/0401048}


\bibitem[Korf et~al\mbox{.}(2010)]%
        {rectpack}
\bibfield{author}{\bibinfo{person}{Richard~E Korf}, \bibinfo{person}{Michael~D Moffitt}, {and} \bibinfo{person}{Martha~E Pollack}.} \bibinfo{year}{2010}\natexlab{}.
\newblock \showarticletitle{Optimal rectangle packing}.
\newblock \bibinfo{journal}{\emph{Annals of Operations Research}}  \bibinfo{volume}{179} (\bibinfo{year}{2010}), \bibinfo{pages}{261--295}.
\newblock


\bibitem[Lamprakos et~al\mbox{.}(2023a)]%
        {idealloc_1}
\bibfield{author}{\bibinfo{person}{Christos~Panagiotis Lamprakos}, \bibinfo{person}{Sotirios Xydis}, \bibinfo{person}{Francky Catthoor}, {and} \bibinfo{person}{Dimitrios Soudris}.} \bibinfo{year}{2023}\natexlab{a}.
\newblock \showarticletitle{The Unexpected Efficiency of Bin Packing Algorithms for Dynamic Storage Allocation in the Wild: An Intellectual Abstract}. In \bibinfo{booktitle}{\emph{Proceedings of the 2023 ACM SIGPLAN International Symposium on Memory Management}} (Orlando, FL, USA) \emph{(\bibinfo{series}{ISMM 2023})}. \bibinfo{publisher}{Association for Computing Machinery}, \bibinfo{address}{New York, NY, USA}, \bibinfo{pages}{58–70}.
\newblock
\showISBNx{9798400701795}
\urldef\tempurl%
\url{https://doi.org/10.1145/3591195.3595279}
\showDOI{\tempurl}


\bibitem[Lamprakos et~al\mbox{.}(2023b)]%
        {beyondrss}
\bibfield{author}{\bibinfo{person}{Christos~Panagiotis Lamprakos}, \bibinfo{person}{Sotirios Xydis}, \bibinfo{person}{Peter Kourzanov}, \bibinfo{person}{Manu Perumkunnil}, \bibinfo{person}{Francky Catthoor}, {and} \bibinfo{person}{Dimitrios Soudris}.} \bibinfo{year}{2023}\natexlab{b}.
\newblock \showarticletitle{Beyond RSS: Towards Intelligent Dynamic Memory Management (Work in Progress)}. In \bibinfo{booktitle}{\emph{Proceedings of the 20th ACM SIGPLAN International Conference on Managed Programming Languages and Runtimes}} (Cascais, Portugal) \emph{(\bibinfo{series}{MPLR 2023})}. \bibinfo{publisher}{Association for Computing Machinery}, \bibinfo{address}{New York, NY, USA}, \bibinfo{pages}{158–164}.
\newblock
\showISBNx{9798400703805}
\urldef\tempurl%
\url{https://doi.org/10.1145/3617651.3622989}
\showDOI{\tempurl}


\bibitem[Lamprou et~al\mbox{.}(2023)]%
        {somas}
\bibfield{author}{\bibinfo{person}{Ioannis Lamprou}, \bibinfo{person}{Zhen Zhang}, \bibinfo{person}{Javier de Juan}, \bibinfo{person}{Hang Yang}, \bibinfo{person}{Yongqiang Lai}, \bibinfo{person}{Etienne Filhol}, {and} \bibinfo{person}{Cedric Bastoul}.} \bibinfo{year}{2023}\natexlab{}.
\newblock \showarticletitle{Safe Optimized Static Memory Allocation for Parallel Deep Learning}. In \bibinfo{booktitle}{\emph{Proceedings of Machine Learning and Systems}}, \bibfield{editor}{\bibinfo{person}{D.~Song}, \bibinfo{person}{M.~Carbin}, {and} \bibinfo{person}{T.~Chen}} (Eds.), Vol.~\bibinfo{volume}{5}. \bibinfo{publisher}{Curan}, \bibinfo{pages}{305--324}.
\newblock
\urldef\tempurl%
\url{https://proceedings.mlsys.org/paper_files/paper/2023/file/676d8419c61f299feb88c28b40edd3b1-Paper-mlsys2023.pdf}
\showURL{%
\tempurl}


\bibitem[Levental(2022)]%
        {levental2022memory}
\bibfield{author}{\bibinfo{person}{Maksim Levental}.} \bibinfo{year}{2022}\natexlab{}.
\newblock \showarticletitle{Memory planning for deep neural networks}.
\newblock \bibinfo{journal}{\emph{arXiv preprint arXiv:2203.00448}} (\bibinfo{year}{2022}).
\newblock


\bibitem[Maas et~al\mbox{.}(2020)]%
        {learningmalloc}
\bibfield{author}{\bibinfo{person}{Martin Maas}, \bibinfo{person}{David~G. Andersen}, \bibinfo{person}{Michael Isard}, \bibinfo{person}{Mohammad~Mahdi Javanmard}, \bibinfo{person}{Kathryn~S. McKinley}, {and} \bibinfo{person}{Colin Raffel}.} \bibinfo{year}{2020}\natexlab{}.
\newblock \showarticletitle{Learning-based Memory Allocation for C++ Server Workloads}. In \bibinfo{booktitle}{\emph{Proceedings of the Twenty-Fifth International Conference on Architectural Support for Programming Languages and Operating Systems}} (Lausanne, Switzerland) \emph{(\bibinfo{series}{ASPLOS '20})}. \bibinfo{publisher}{Association for Computing Machinery}, \bibinfo{address}{New York, NY, USA}, \bibinfo{pages}{541–556}.
\newblock
\showISBNx{9781450371025}
\urldef\tempurl%
\url{https://doi.org/10.1145/3373376.3378525}
\showDOI{\tempurl}


\bibitem[Maas et~al\mbox{.}(2022)]%
        {telamalloc}
\bibfield{author}{\bibinfo{person}{Martin Maas}, \bibinfo{person}{Ulysse Beaugnon}, \bibinfo{person}{Arun Chauhan}, {and} \bibinfo{person}{Berkin Ilbeyi}.} \bibinfo{year}{2022}\natexlab{}.
\newblock \showarticletitle{TelaMalloc: Efficient On-Chip Memory Allocation for Production Machine Learning Accelerators}. In \bibinfo{booktitle}{\emph{Proceedings of the 28th ACM International Conference on Architectural Support for Programming Languages and Operating Systems, Volume 1}} (Vancouver, BC, Canada) \emph{(\bibinfo{series}{ASPLOS 2023})}. \bibinfo{publisher}{Association for Computing Machinery}, \bibinfo{address}{New York, NY, USA}, \bibinfo{pages}{123–137}.
\newblock
\showISBNx{9781450399159}
\urldef\tempurl%
\url{https://doi.org/10.1145/3567955.3567961}
\showDOI{\tempurl}


\bibitem[Moffitt(2024)]%
        {minimalloc}
\bibfield{author}{\bibinfo{person}{Michael~D. Moffitt}.} \bibinfo{year}{2024}\natexlab{}.
\newblock \showarticletitle{MiniMalloc: A Lightweight Memory Allocator for Hardware-Accelerated Machine Learning}. In \bibinfo{booktitle}{\emph{Proceedings of the 28th ACM International Conference on Architectural Support for Programming Languages and Operating Systems, Volume 4}} (Vancouver, BC, Canada) \emph{(\bibinfo{series}{ASPLOS '23})}. \bibinfo{publisher}{Association for Computing Machinery}, \bibinfo{address}{New York, NY, USA}, \bibinfo{pages}{238–252}.
\newblock
\showISBNx{9798400703942}
\urldef\tempurl%
\url{https://doi.org/10.1145/3623278.3624752}
\showDOI{\tempurl}


\bibitem[Navasca et~al\mbox{.}(2023)]%
        {navasca}
\bibfield{author}{\bibinfo{person}{Christian Navasca}, \bibinfo{person}{Martin Maas}, \bibinfo{person}{Petros Maniatis}, \bibinfo{person}{Hyeontaek Lim}, {and} \bibinfo{person}{Guoqing~Harry Xu}.} \bibinfo{year}{2023}\natexlab{}.
\newblock \showarticletitle{Predicting Dynamic Properties of Heap Allocations using Neural Networks Trained on Static Code: An Intellectual Abstract}. In \bibinfo{booktitle}{\emph{Proceedings of the 2023 ACM SIGPLAN International Symposium on Memory Management}} (Orlando, FL, USA) \emph{(\bibinfo{series}{ISMM 2023})}. \bibinfo{publisher}{Association for Computing Machinery}, \bibinfo{address}{New York, NY, USA}, \bibinfo{pages}{43–57}.
\newblock
\showISBNx{9798400701795}
\urldef\tempurl%
\url{https://doi.org/10.1145/3591195.3595275}
\showDOI{\tempurl}


\bibitem[Oh et~al\mbox{.}(2022)]%
        {maphea}
\bibfield{author}{\bibinfo{person}{Deok-Jae Oh}, \bibinfo{person}{Yaebin Moon}, \bibinfo{person}{Do~Kyu Ham}, \bibinfo{person}{Tae~Jun Ham}, \bibinfo{person}{Yongjun Park}, \bibinfo{person}{Jae~W. Lee}, \bibinfo{person}{Jung~Ho Ahn}, {and} \bibinfo{person}{Eojin Lee}.} \bibinfo{year}{2022}\natexlab{}.
\newblock \showarticletitle{MaPHeA: A Framework for Lightweight Memory Hierarchy-aware Profile-guided Heap Allocation}.
\newblock \bibinfo{journal}{\emph{ACM Trans. Embed. Comput. Syst.}} \bibinfo{volume}{22}, \bibinfo{number}{1}, Article \bibinfo{articleno}{2} (\bibinfo{date}{Dec.} \bibinfo{year}{2022}), \bibinfo{numpages}{28}~pages.
\newblock
\showISSN{1539-9087}
\urldef\tempurl%
\url{https://doi.org/10.1145/3527853}
\showDOI{\tempurl}


\bibitem[Pisarchyk and Lee(2020)]%
        {pisarchyk2020efficient}
\bibfield{author}{\bibinfo{person}{Yury Pisarchyk} {and} \bibinfo{person}{Juhyun Lee}.} \bibinfo{year}{2020}\natexlab{}.
\newblock \showarticletitle{Efficient memory management for deep neural net inference}.
\newblock \bibinfo{journal}{\emph{arXiv preprint arXiv:2001.03288}} (\bibinfo{year}{2020}).
\newblock


\bibitem[Robson(1971)]%
        {robson_1}
\bibfield{author}{\bibinfo{person}{J.~M. Robson}.} \bibinfo{year}{1971}\natexlab{}.
\newblock \showarticletitle{An Estimate of the Store Size Necessary for Dynamic Storage Allocation}.
\newblock \bibinfo{journal}{\emph{J. ACM}} \bibinfo{volume}{18}, \bibinfo{number}{3} (\bibinfo{date}{July} \bibinfo{year}{1971}), \bibinfo{pages}{416–423}.
\newblock
\showISSN{0004-5411}
\urldef\tempurl%
\url{https://doi.org/10.1145/321650.321658}
\showDOI{\tempurl}


\bibitem[Robson(1974)]%
        {robson_2}
\bibfield{author}{\bibinfo{person}{J.~M. Robson}.} \bibinfo{year}{1974}\natexlab{}.
\newblock \showarticletitle{Bounds for Some Functions Concerning Dynamic Storage Allocation}.
\newblock \bibinfo{journal}{\emph{J. ACM}} \bibinfo{volume}{21}, \bibinfo{number}{3} (\bibinfo{date}{July} \bibinfo{year}{1974}), \bibinfo{pages}{491–499}.
\newblock
\showISSN{0004-5411}
\urldef\tempurl%
\url{https://doi.org/10.1145/321832.321846}
\showDOI{\tempurl}


\bibitem[Savage and Jones(2020)]%
        {halo}
\bibfield{author}{\bibinfo{person}{Joe Savage} {and} \bibinfo{person}{Timothy~M. Jones}.} \bibinfo{year}{2020}\natexlab{}.
\newblock \showarticletitle{HALO: post-link heap-layout optimisation}. In \bibinfo{booktitle}{\emph{Proceedings of the 18th ACM/IEEE International Symposium on Code Generation and Optimization}} (San Diego, CA, USA) \emph{(\bibinfo{series}{CGO '20})}. \bibinfo{publisher}{Association for Computing Machinery}, \bibinfo{address}{New York, NY, USA}, \bibinfo{pages}{94–106}.
\newblock
\showISBNx{9781450370479}
\urldef\tempurl%
\url{https://doi.org/10.1145/3368826.3377914}
\showDOI{\tempurl}


\bibitem[Scherer et~al\mbox{.}(2024)]%
        {deeploy}
\bibfield{author}{\bibinfo{person}{Moritz Scherer}, \bibinfo{person}{Luka Macan}, \bibinfo{person}{Victor J.~B. Jung}, \bibinfo{person}{Philip Wiese}, \bibinfo{person}{Luca Bompani}, \bibinfo{person}{Alessio Burrello}, \bibinfo{person}{Francesco Conti}, {and} \bibinfo{person}{Luca Benini}.} \bibinfo{year}{2024}\natexlab{}.
\newblock \showarticletitle{Deeploy: Enabling Energy-Efficient Deployment of Small Language Models on Heterogeneous Microcontrollers}.
\newblock \bibinfo{journal}{\emph{IEEE Transactions on Computer-Aided Design of Integrated Circuits and Systems}} \bibinfo{volume}{43}, \bibinfo{number}{11} (\bibinfo{year}{2024}), \bibinfo{pages}{4009--4020}.
\newblock
\urldef\tempurl%
\url{https://doi.org/10.1109/TCAD.2024.3443718}
\showDOI{\tempurl}


\bibitem[Sekiyama et~al\mbox{.}(2018)]%
        {sekiyama2018profileguidedmemoryoptimizationdeep}
\bibfield{author}{\bibinfo{person}{Taro Sekiyama}, \bibinfo{person}{Takashi Imamichi}, \bibinfo{person}{Haruki Imai}, {and} \bibinfo{person}{Rudy Raymond}.} \bibinfo{year}{2018}\natexlab{}.
\newblock \bibinfo{title}{Profile-guided memory optimization for deep neural networks}.
\newblock
\newblock
\showeprint[arxiv]{1804.10001}~[cs.DC]
\urldef\tempurl%
\url{https://arxiv.org/abs/1804.10001}
\showURL{%
\tempurl}


\bibitem[Tillet et~al\mbox{.}(2019)]%
        {triton}
\bibfield{author}{\bibinfo{person}{Philippe Tillet}, \bibinfo{person}{H.~T. Kung}, {and} \bibinfo{person}{David Cox}.} \bibinfo{year}{2019}\natexlab{}.
\newblock \showarticletitle{Triton: an intermediate language and compiler for tiled neural network computations}. In \bibinfo{booktitle}{\emph{Proceedings of the 3rd ACM SIGPLAN International Workshop on Machine Learning and Programming Languages}} (Phoenix, AZ, USA) \emph{(\bibinfo{series}{MAPL 2019})}. \bibinfo{publisher}{Association for Computing Machinery}, \bibinfo{address}{New York, NY, USA}, \bibinfo{pages}{10–19}.
\newblock
\showISBNx{9781450367196}
\urldef\tempurl%
\url{https://doi.org/10.1145/3315508.3329973}
\showDOI{\tempurl}


\bibitem[Wilson et~al\mbox{.}(1995)]%
        {wilson}
\bibfield{author}{\bibinfo{person}{Paul~R. Wilson}, \bibinfo{person}{Mark~S. Johnstone}, \bibinfo{person}{Michael Neely}, {and} \bibinfo{person}{David Boles}.} \bibinfo{year}{1995}\natexlab{}.
\newblock \showarticletitle{Dynamic storage allocation: A survey and critical review}. In \bibinfo{booktitle}{\emph{Memory Management}}, \bibfield{editor}{\bibinfo{person}{Henry~G. Baler}} (Ed.). \bibinfo{publisher}{Springer Berlin Heidelberg}, \bibinfo{address}{Berlin, Heidelberg}, \bibinfo{pages}{1--116}.
\newblock
\showISBNx{978-3-540-45511-0}


\bibitem[Zhao et~al\mbox{.}(2025)]%
        {memoLLM}
\bibfield{author}{\bibinfo{person}{Pinxue Zhao}, \bibinfo{person}{Hailin Zhang}, \bibinfo{person}{Fangcheng Fu}, \bibinfo{person}{Xiaonan Nie}, \bibinfo{person}{Qibin Liu}, \bibinfo{person}{Fang Yang}, \bibinfo{person}{Yuanbo Peng}, \bibinfo{person}{Dian Jiao}, \bibinfo{person}{Shuaipeng Li}, \bibinfo{person}{Jinbao Xue}, \bibinfo{person}{Yangyu Tao}, {and} \bibinfo{person}{Bin Cui}.} \bibinfo{year}{2025}\natexlab{}.
\newblock \showarticletitle{MEMO: Fine-grained Tensor Management For Ultra-long Context LLM Training}.
\newblock \bibinfo{journal}{\emph{Proc. ACM Manag. Data}} \bibinfo{volume}{3}, \bibinfo{number}{1}, Article \bibinfo{articleno}{53} (\bibinfo{date}{Feb.} \bibinfo{year}{2025}), \bibinfo{numpages}{28}~pages.
\newblock
\urldef\tempurl%
\url{https://doi.org/10.1145/3709703}
\showDOI{\tempurl}


\end{thebibliography}

\section*{APPENDIX}
\appendix
\section{Lemma 1}
What follows is a verbatim copy of Lemma 1 by Buchsbaum et al.~\cite{buchsbaum}. As in the main text, we represent parts of the proof that are of no concern to implementing the FU with ``$[...]$''.

\begin{l1}
Given a set $Y$ of unit-height jobs, all live at some fixed x-coordinate $t$, an integer box-height parameter $H$, and a sufficiently small $\epsilon > 0$, there exist a subset $Y'$ of $Y$, $|Y-Y'|~\leq~2H\lceil1/\epsilon^2\rceil$, a set $B$ of boxes, each of height $H$, and a boxing of $Y'$ into $B$ such that at any x-coordinate $u$,

\begin{equation*}
    L_B(u)~\leq~L_{Y'}(u)~+~4\epsilon L_Y(u)
\end{equation*}
\end{l1}

\begin{proof}
    $[...]$ partition the jobs of $Y$ into \textit{strips} $[...]$. The first two strips are defined as follows.

    \begin{itemize}
        \item Create a vertical strip consisting of the $H\lceil1/\epsilon^2\rceil$ jobs with the earliest starting x-coordinates (or fewer if there are not enough jobs)
        \item If any jobs remain, create a horizontal strip consisting of the $H\lceil1/\epsilon^2\rceil$ jobs that remain with the latest ending x-coordinates (or fewer if not enough jobs remain)
    \end{itemize}

    Define $Y'$ to be the set of all jobs \textit{not} in the first vertical or first horizontal strip. $[...]$ Now partition the jobs of $Y'$ as follows. As long as there are jobs remaining, repeat the following.

    \begin{itemize}
        \item Create a vertical strip consisting of the $H\lceil1/\epsilon\rceil$ jobs that remain with the earliest starting x-coordinates (or fewer if there are not enough jobs left)
        \item If any jobs remain, create a horizontal strip consisting of the $H\lceil1/\epsilon\rceil$ jobs that remain with the latest ending x-coordinates (or fewer if not enough jobs remain)
    \end{itemize}

    Now, for every vertical strip of $Y'$, take the jobs in order of decreasing ending x-coordinate, in groups of size $H$ (the last group of the last strip possibly smaller), and box them. Similarly, for every horizontal strip of $Y'$, take the jobs in order of increasing starting x-coordinate, in groups of size $H$ (the last group of the last strip possibly smaller), and box them. $[...]$
\end{proof}

We call the jobs in $Y-Y'$ \textit{unresolved jobs}.

\section{The Impossibility of Theorem 19}
As above, we begin with a verbatim copy of Theorem 19~\cite{buchsbaum}:

\begin{t19}
For all $\epsilon>0$, there exists a polynomial-time $(2+\epsilon)$-approximation algorithm for DSA.
\end{t19}

\begin{proof}
    Consider some small positive $\delta$ to be determined. Let $X=X_s~\cup~X_l$, where $X_s$ is the set of jobs of height less than $\delta^7L$ and $X_l=X~\setminus~X_s$. Use Theorem 16 with error parameter $\delta$ to pack the jobs in $X_s$, yielding a $(1+c'\delta)$-approximation for some constant $c'$. Apply the $(1+\delta)$-approximation algorithm implied by Theorem 12 with the same $\delta$ to pack the jobs in $X_l$, which is possible because the load divided by the minimum height is at most $1/\delta^7$, which is certainly at most $C~log_2n/log_2log_2n$ for $C=1/\delta^7$; this yields a $(1+\delta)$ approximation. Choose $\delta$ so that $\delta(c'+1)=\epsilon$.
\end{proof}

The impossibility of writing the above as a computer program function is evident. The parameter $\delta$ governs all steps, but is only determined in the end. Nevertheless, given the liberties we have taken with the rest of BA in order to make it functional, future attempts to ``hack'' Theorem 19 might prove fruitful.

\end{document}